\documentclass[useAMS,usenatbib,usegraphicx]{mn2e}
\usepackage{epsf}

 \newcommand{\bm}[1]{\mbox{\boldmath$#1$}}

\newcommand{\simgt}{\lower.5ex\hbox{$\; \buildrel > \over \sim \;$}}
\newcommand{\simlt}{\lower.5ex\hbox{$\; \buildrel < \over \sim \;$}}

\title[Direct measurement of dark matter halo ellipticity]
{Direct measurement of dark matter halo ellipticity from
two-dimensional lensing shear maps of 25 massive clusters
\thanks{Based on
data collected at Subaru Telescope and obtained from the SMOKA, which
is operated by the Astronomy Data Center, National Astronomical
Observatory of Japan.}  
}

\author[M.~Oguri et al.]
{Masamune Oguri,$^{1,2}$\thanks{E-mail: masamune.oguri@nao.ac.jp} 
Masahiro Takada,$^{3}$
Nobuhiro Okabe$^{4,5}$
and Graham P. Smith$^{6}$\\
$^1$Division of Theoretical Astronomy, National Astronomical
Observatory of Japan, 2-21-1 Osawa, Mitaka, Tokyo 181-8588, Japan.\\ 
$^2$Kavli Institute for Particle Astrophysics and Cosmology, 
Stanford University, 2575 Sand Hill Road, Menlo Park, CA
94025, USA.\\
$^3$Institute for the Physics and Mathematics of the Universe
  (IPMU), The University of Tokyo, Chiba 277-8582, Japan.\\
$^4$Institute of Astronomy and Astrophysics, Academica Sinica, PO
  Box 23-141, Taipei 106, Taiwan.\\
$^5$Astronomical institute, Tohoku University, Aramaki, Aoba-ku,
  Sendai, 980-8578, Japan.\\
$^6$School of Physics and Astronomy, University of Birmingham, Edgbaston,
Birmingham B15 2TT.\\
} 

\begin{document}

\date{\today}

\voffset- .5in

\pagerange{\pageref{firstpage}--\pageref{lastpage}} \pubyear{}

\maketitle

\label{firstpage}

\begin{abstract}
We present new measurements of dark matter distributions in 25 X-ray
luminous clusters by making a full use of the two-dimensional (2D)
weak lensing signals obtained from high-quality Subaru/Suprime-Cam
imaging data. Our approach to directly compare the measured lensing
shear pattern with elliptical model predictions allows us to extract
new information on the mass distributions of individual clusters, such
as the halo ellipticity and mass centroid. We find that these
parameters on the cluster shape are little degenerate with cluster
mass and concentration parameters. By combining the 2D fitting results
for a subsample of 18 clusters, the elliptical shape of dark matter
haloes is detected at $7\sigma$ significance level. The mean
ellipticity is found to be 
$\langle e \rangle = \langle 1-b/a \rangle =0.46\pm 0.04$ ($1\sigma$), 
which is in excellent agreement with a theoretical prediction based on
the standard collisionless cold dark matter model. The mass centroid
can be constrained with a typical accuracy of  $\sim 20$~arcseconds
($\sim 50h^{-1}$~kpc) in radius for each cluster. The mass centroid
position fairly well matches the position of the brightest cluster galaxy,
with some clusters showing significant offsets. Thus the 2D shear
fitting method enables to assess one of the most important systematic
errors inherent in the stacked cluster weak lensing technique, the mass
centroid uncertainty.  In addition, the shape of the dark mass
distribution is found to be only weakly correlated with that of the
member galaxy distribution or the brightest cluster galaxy. We
carefully examine possible sources of systematic errors in our
measurements including the effect of substructures, the cosmic shear
contamination, fitting regions, and the dilution effect, and find none
of them to be significant. Our results demonstrate the power of
high-quality imaging data for exploring the detailed spatial
distribution of dark matter, which should improve the ability of
future surveys to conduct cluster cosmology experiments.   
\end{abstract}


\section{Introduction}
\label{sec:intro}

The internal structure of clusters predicted by $N$-body simulations
of the collisionless Cold Dark Matter (CDM) universe exhibits several
important features, including the universal radial density profile
with progressively shallower slopes toward the centre  
\citep[e.g.,][]{navarro96,navarro97} and the highly non-spherical
structure fitted well by a triaxial density profile
\citep[e.g.,][]{jing02}. These features are closely related to the
cold and collisionless nature of hypothetical dark matter particles
\citep[e.g.,][]{ostriker03}, which suggests that detailed
observational tests of cluster density profiles may provide a clue to
the nature of dark mater. In addition, an accurate determination of
the mass distribution of individual clusters is also important in
calibrating observable-mass relations which are necessary for
constraining cosmological parameters from cluster abundance 
\citep[e.g.,][]{bahcall93,eke96,kitayama97,henry00,borgani01,
bahcall03,dahle06,mantz08,henry09,vikhlinin09,rozo10}. 

Gravitational lensing is one of the most powerful methods to constrain 
the cluster mass distribution. This is because gravitational lensing
probes the matter distribution directly, regardless of whether it is
luminous or dark. This is in marked contrast with other probes such as
X-ray and Sunyaev-Zel'dovich effect for which assumptions on gas state
and other pressure components are crucial in extracting the mass
distribution. In particular the cluster mass distribution has
extensively been studied using weak lensing technique which makes use
of small distortions of background galaxies produced by a foreground
cluster \citep[see][for a review]{bartelmann01}. Indeed the weak lens
technique has been quite successful in constraining dark matter
distributions in clusters independently of the distribution of the hot
gas component \citep{kneib03,clowe06,jee07,mahdavi07,mahdavi08,okabe08,bradac08}. 

While the weak lensing technique allows the non-parametric
reconstruction of the mass map \citep{kaiser93}, azimuthally-averaged 
one-dimensional (1D) tangential shear profiles have been used in most
of quantitative studies so far
\citep[e.g.,][]{bardeau07,broadhurst08,hamana09,oguri09b,okabe10}.  
The main reason for this is that the 1D profile fitting provides an
easy and efficient way to extract several important cluster
properties, such as mass, concentration, and radial density profiles
of clusters. However, the highly non-spherical nature of CDM haloes
predicted by $N$-body simulations indicates that the full
two-dimensional (2D) study of weak lensing data is crucial for
further tests of the cluster mass distribution. Moreover, an important 
disadvantage of the 1D profile fitting is that we have to assume the
centre of the cluster a priori. It has often been assumed that the
mass centroid coincides with the location of the brightest cluster
galaxy (BCG), despite the fact that offsets of BCGs from the mass
centroids appear to be common 
\citep[e.g.,][]{katayama03,lin04,koester07,ho09,sanderson09}.  The 2D
weak lensing 
studies with non-spherical mass models have been attempted, but only
for a handful of lensing clusters \citep[e.g.,][]{cypriano04,deb10}.
Fitting of weak lensing data by taking fully into account a 3D
(triaxial) cluster shape and the projection along arbitrary
line-of-sight has been performed by \citet{oguri05}, and later
by \citet{corless09}, but their interests lied in the determination of
masses and concentration parameters rather than the shape of the
projected mass distribution. Given the strong selection bias of the
apparent 2D cluster ellipticity for lensing clusters \citep{oguri09a},
more systematic 2D weak lensing studies of massive clusters should be
conducted using well-defined statistical cluster samples.    

Cluster mass distributions have also been studied in details using
stacked weak lensing data \citep{sheldon07a,sheldon07b,johnston07,
mandelbaum06b,mandelbaum08}. Although the stacked weak lensing
technique is quite successful in extracting average properties of
clusters, its main disadvantage comes from the centroid problem
\citep[e.g.,][]{johnston07,mandelbaum10}; in stacking clusters one has
to assume a priori the mass centroid for each cluster. The
misidentification of the mass centroid results in the suppression of
weak lensing signals, particularly near the cluster centre. Recently,
\citet{evans09} measured the ellipticity of isolated clusters with
stacked weak lensing technique to obtain the average cluster
ellipticity of $e\sim 0.5$.  However, their result is built on the
assumption that the cluster mass distributions is aligned with the
spatial distribution of member galaxies, and also on the assumption
about the cluster centre, which  makes the interpretation somewhat
difficult. 

In this paper, we present a systematic 2D study of weak lensing
maps. We analyse 25 clusters at $0.15<z<0.3$ presented by
\citet{okabe10}, who examined azimuthally-averaged 1D tangential
profiles to study the radial mass profiles of the clusters.  The
cluster sample is essentially X-ray flux-limited statistical sample
from the Local Cluster Substructure Survey (LoCuSS; G. Smith et al.,
in preparation). In this paper we extend the lensing analysis of 
\citet{okabe10} and perform a full 2D fitting of the shear fields with
particular emphasis on the projected 2D ellipticities of the
clusters. All the weak lensing data are based on optical images taken
with Subaru/Suprime-cam \citep{miyazaki02} which is known to be the
current best ground facility for weak lensing studies because of its
exquisite image quality and wide field-of-view
\citep[e.g.,][]{kasliwal08}. In our analysis we leave the centre of
the mass profile as a free parameter and fit it simultaneously in
order to see how the 2D analysis of shear maps constrains the cluster
mass centroid which has often been assumed to coincide with the
location of the BCG. Our approach to fit individual clusters with the
centroid and orientation of the cluster as free parameters not only
overcomes the problems inherent to stacked weak lensing technique but
also allows us to even test the assumptions on the centroid and
orientation from weak lensing data alone.  

The structure of this paper is as follows. We describe our fitting
procedure in \S\ref{sec:fit}, and present results in
\S\ref{sec:result}. In \S\ref{sec:discussion} we discuss possible
systematic effects, and in \S\ref{sec:conclusion} we draw our
conclusions. Throughout the paper we assume a flat universe with the
matter density $\Omega_M=1-\Omega_\Lambda=0.274$, the dimensionless
Hubble constant $h=0.704$, the baryon density $\Omega_b h^2=0.02267$,
the spectral index $n_s=0.96$, and the normalisation of the matter
power spectrum $\sigma_8=0.812$ \citep{komatsu09}. 

\section{Data Analysis}
\label{sec:fit}

\subsection{Cluster sample}

In this paper, we study the same cluster sample as in \citet{okabe10}.
The sample comprises 30 massive clusters observed with
Subaru/Suprime-cam, and represents an unbiased subsample of the LoCuSS
clusters, an all-sky sample of $\sim 100$ X-ray luminous ($L_X\ga
2\times10^{44}{\rm erg\,s^{-1}}$) galaxy clusters located at
$0.15<z<0.3$ selected from the ROSAT All Sky Survey catalogues
\citep{ebeling98,ebeling00,bohringer04}. Since X-ray properties of
clusters are not very much affected by the degree of the elongation
along the line-of-sight \citep[e.g.,][]{gavazzi05}, the orientation
bias, which is important for lensing-selected clusters
\citep{hennawi07,oguri09a}, is expected to be insignificant for our
sample. 

The properties of the clusters and the basic parameters of the
Subaru/Suprime-cam observations are summarised in Table~1 of
\citet{okabe10}. All the 30 clusters were observed with sub-arcsecond
seeing conditions. Among these 25 clusters have two-filter images
(mostly $V$- and $i$-bands, with typical exposure time of $\sim
20-40$~minutes for each filter), whereas for the other 5 clusters only
one-filter images are available. The colour information is very
important to minimise the dilution effect by cluster member galaxies
\citep[see also][]{broadhurst05}. In addition, the colour information
allows to define a sample of member galaxies from galaxies around the
red sequence locus in the colour-magnitude diagram, and to compare the
member galaxy distribution with the mass distribution constrained from
lensing observables. Thus, in this paper we restrict our analysis to
the 25 clusters with colour information. As a fiducial background
galaxy population, we adopt ``red+blue'' galaxy sample defined as
faint galaxies with colours redder or bluer than the red-sequence by
properly chosen offsets. As shown in \citet{okabe10}, the dilution
effect by member galaxies appears to be quite small for this source
galaxy sample. The effective source redshift, which is defined such
that the lensing depth $D_{ls}/D_s$ ($D_{ls}$ and $D_s$ being the
angular diameter distances from the lens to the source and from the
observer to the source, respectively) at that redshift becomes equal to
the mean lensing depth averaged over the background galaxy population, 
is estimated by matching  magnitudes and colours of the background
galaxies to COSMOS photometric redshift catalogue of \citet{ilbert09}. 
See \citet{okabe10} for more detailed descriptions.

\begin{figure*}
\begin{center}
 \includegraphics[width=0.47\hsize]{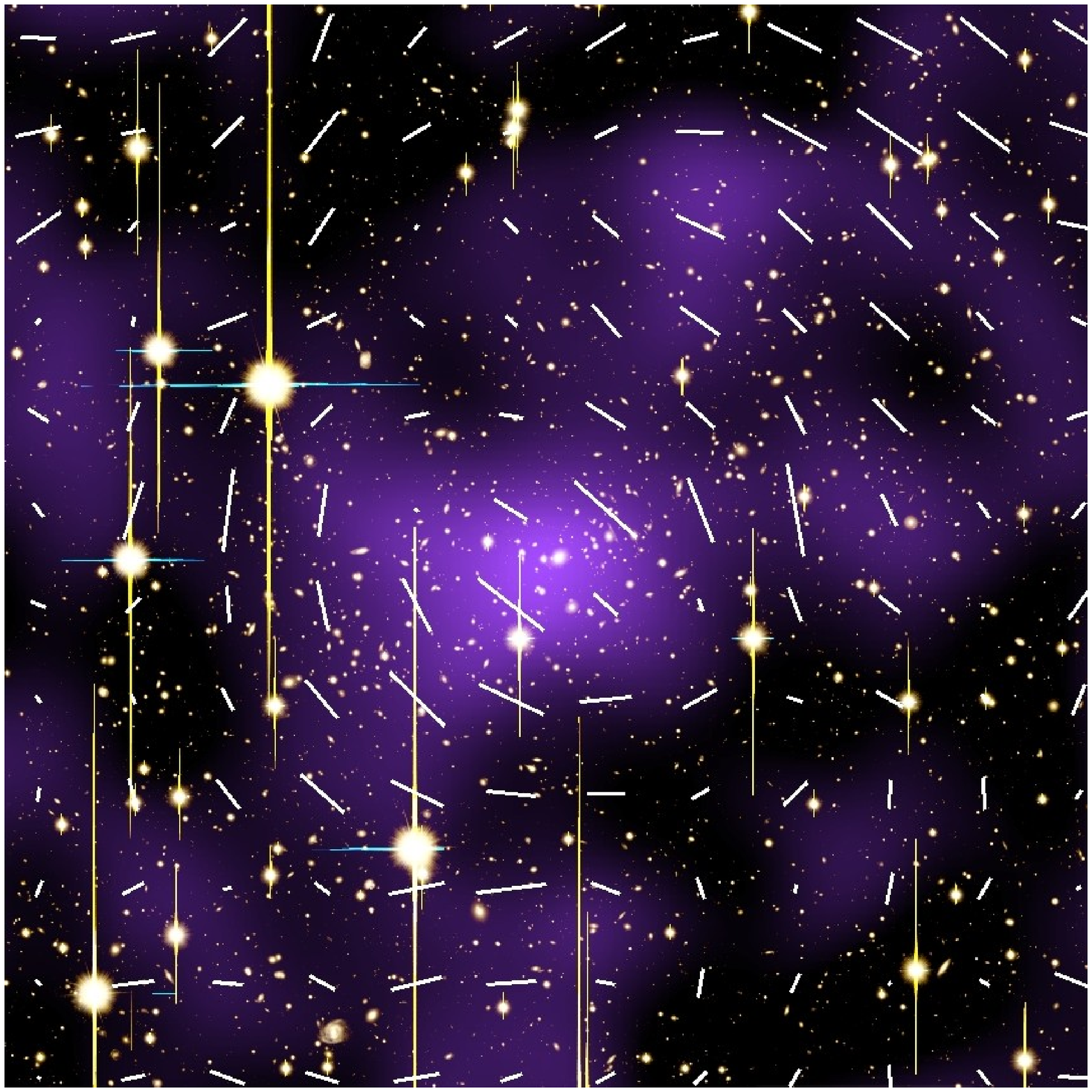}
 \includegraphics[width=0.49\hsize]{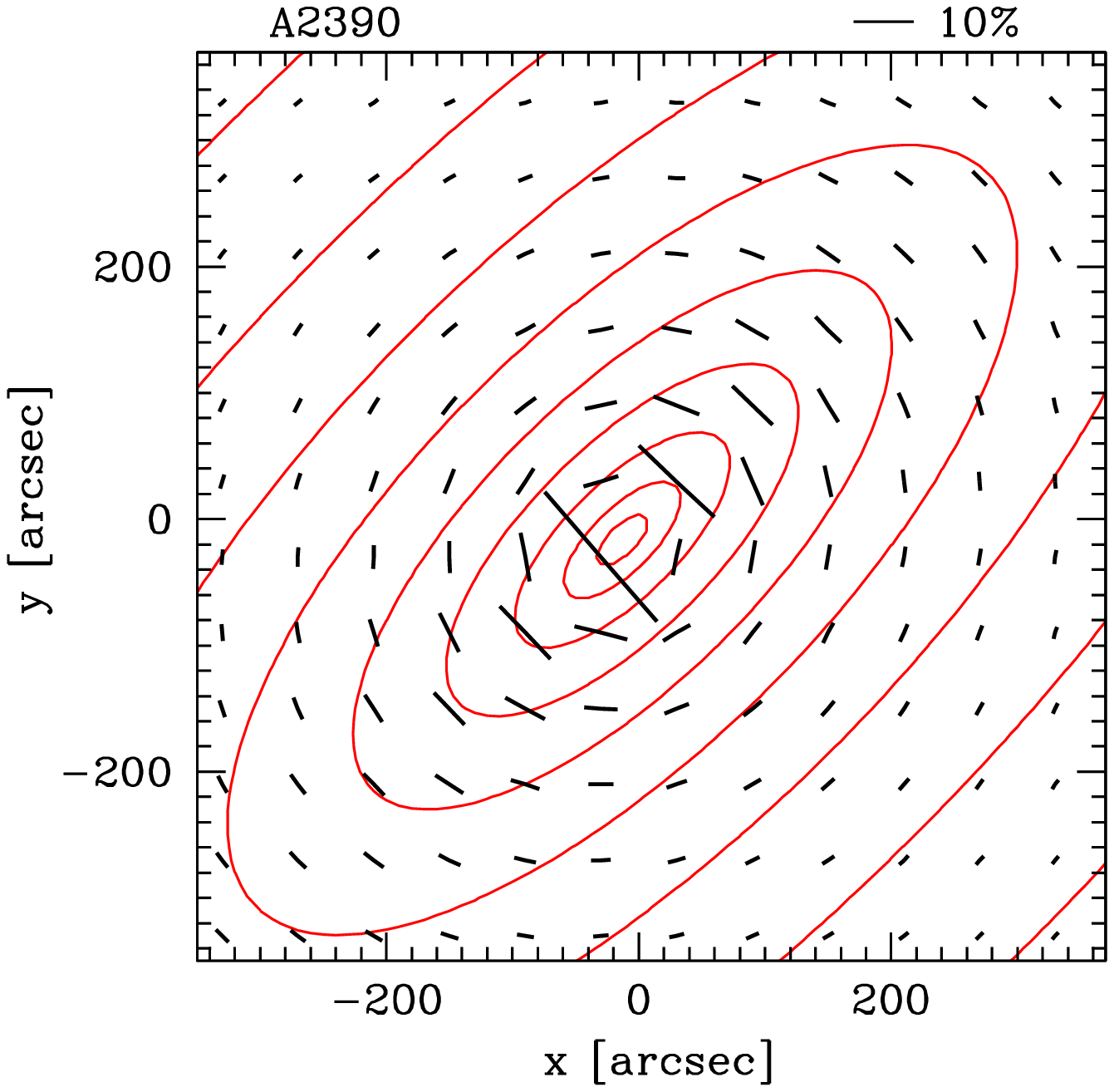}
\end{center}
\caption{{\it Left panel:} An example of our weak lensing measurement for
 A2390. The size of Each panel is $12'\times 12'$. The stick in each
 $1'\times' 1$ pixel shows the distortion field estimated from
 background galaxy images contained within the pixel, where a
 background galaxy image is deformed along the stick direction, and
 the length is proportional to the shear amplitude. The shear field in
 this panel is smoothed with a Gaussian with the full width at half
 maximum of $\simeq 1.6'$ for illustrative purpose. Overplotted is
 the surface mass density map reconstructed from the weak lensing shear
 measurement \citep[see][]{okabe10}. North is up and East is
 left. {\it Right panel:}  The shear field predicted by our best-fit 
 elliptical NFW model (see also Figure~\ref{fig:cont_a2390} and
Table~\ref{table:fit}), while the contours are the isodensity map. The
 best-fit ellipticity of the projected mass density is 
$e\equiv 1-b/a=0.598$. 
\label{fig:smap_a2390}}
\end{figure*}

\subsection{Weak lensing data}
\label{sec:wldata}

Weak lensing distortion measurements enable a direct reconstruction of
the projected 2D mass distribution \citep{kaiser93}. However, the
non-local nature of the mass reconstruction indicates that
reconstructed mass densities between different pixels in the mass map
are highly correlated. Thus one has to take account of the full
covariance matrix of the mass map in order to extract proper
information from the mass map, as done in \citet{oguri05} and
\citet{umetsu08}. In this paper, we avoid this complication by 
working directly on the 2D distortion maps.  

We use the weak lensing distortion measurements presented by
\citet{okabe10}. For each cluster field, the reduced distortion, 
$g_\alpha=\gamma_\alpha/(1-\kappa)$, was estimated from the shape of
each source galaxy by analysing the Subaru/Suprime-cam images based on 
the algorithm described in \citet{kaiser95}. Throughout this paper we
employ the convention that the Greek subscripts denote the two
components of distortion, e.g., $\alpha=1$ or $2$. The lensing
distortion is not measurable from individual galaxy image due to
the dominant intrinsic shape noise. Instead the signal is measurable
in a statistical sense, i.e., by averaging the galaxy shapes over a
sufficient number of galaxies. Given the deep Subaru images and the
distortion strength in a cluster region, the angular resolution of weak
lensing distortion is an arcminute scale. In order to conduct the full
two-dimensional analysis of distortion fields, we employ the pixelised  
data of background galaxy shapes. Specifically, the reduced shear in
the $l$-th pixel (its angular position $\bm{\theta}_l$) is estimated
as 
\begin{equation}
  \langle g_\alpha \rangle(\bm{\theta}_l) =\left[
\sum_{\bm{\theta}_i\in \bm{\theta}_l}w_i g_\alpha(\bm{\theta}_i)
\right]\left[\sum_{\bm{\theta}_i\in \bm{\theta}_l} w_i\right]^{-1},
\end{equation}
where $\bm{\theta}_i$ denotes the angular position of the $i$-th source
galaxy and the summation runs over source galaxies contained within 
the $l$-th pixel. Following \citet{okabe10} we use the weighting of
each galaxy shape such that a galaxy whose shape is more reliably
measured is assigned a larger weight. The weight $w_i$ for the $i$-th
galaxy is given by
\begin{equation}
  w_i=\frac{1}{a^2+\sigma_{g(i)}^2},
\end{equation}
with $a=0.4$ and $\sigma_{g(i)}$ being the uncertainty of shape
measurement for each galaxy \citep[see][]{okabe10}. As stated above, the
dominant source of distortion measurement error is the intrinsic galaxy
shape. The shape noise in each pixel is estimated as 
\begin{equation}
  \sigma_g^2(\bm{\theta}_l) =\frac{1}{2}\left[
\sum_{\bm{\theta}_i\in\bm{\theta}_l} w_i^2 \sigma_{g(i)}^2\right]
\left[\sum_{\bm{\theta}_i\in\bm{\theta}_l} w_i\right]^{-2}.
\label{eq:shapenoise}
\end{equation}

For all the clusters we adopt a grid size of $1'\times1'$ (we
tried smaller grid sizes and found that the results are almost
unchanged; see also Appendix~\ref{sec:app}). We do not
use four innermost grids ($2'\times2'$ box) for the fitting, because
source galaxies are obscured by dense distribution of cluster member
galaxies especially in the central region, and also because our
assumption of single source redshift may become inaccurate near the
cluster centre due to the fewer sampling of source galaxies. Moreover, 
the weak lensing approximation breaks down near the cluster centre.
Although the field-of-view of Subaru/Suprime-cam is $34'\times27'$, we
conduct our fitting only in a $20'\times20'$ region ($20'$ corresponds
to physical transverse sizes of $2.2h^{-1}$~Mpc for $z=0.15$ and
$3.8h^{-1}$~Mpc for $z=0.3$, respectively) 
centred at the BCG, which roughly corresponds to virial radii of
clusters in our sample, in order to reduce the projection effect,
i.e. the effect of 
different structures along the same line of sight, which is more
prominent in the boundary region where the cluster lensing signal is
very weak. However, it should be noted that this restriction of
fitting region little affects the final results, as we will discuss
later in more detail. Thus the total number of girds used for the
fitting is $N_{\rm pixel}=396$. An example of our weak lensing shear
map is given in the left panel of Figure~\ref{fig:smap_a2390}.

For the range of angular scales we use for the fitting, the measured
distortion field is nearly the shear field, $g_\alpha\simeq
\gamma_\alpha$, which we will simply assume in the following analysis.

\begin{table*}
 \caption{Results of weak lensing analysis for 25 clusters, obtained
   by fitting the 2D shear maps with predictions of the elliptical NFW
   model.  Note that the degree of freedom of the $\chi^2$ fitting is
   786.  The coordinate origin in the fitting is taken as the BCG
   position for each cluster. The position angle $\theta_e$ is defined
   East of North. Each error indicates the 1$\sigma$ error 
  marginalised over other parameter uncertainties.
\label{table:fit}}   
 \begin{tabular}{@{}ccccccccc}
  \hline
   Name
   & $z$
   & $\chi^2$ 
   & $x_{\rm c}$
   & $y_{\rm c}$
   & $M_{\rm vir}$
   & $c_{\rm vir}$ 
   & $e$ 
   & $\theta_e$ \\
   &
   & 
   & (arcsec) 
   & (arcsec) 
   & ($10^{14}h^{-1}M_\odot$) 
   & 
   & 
   & (deg) \\
 \hline
           A68 & 0.2546 &  719.1 & $  44.8^{ +12.2}_{ -13.2}$ & $  23.7^{ +16.9}_{ -19.6}$ & $  7.83^{ +6.84}_{ -3.95}$ & $  3.10^{ +3.71}_{ -2.04}$ & $ 0.552^{+0.174}_{-0.243}$ & $  -3.3^{  +9.9}_{ -38.1}$ \\ 
          A115$^a$ & 0.1971 &  725.0 & $ -78.5^{ +49.9}_{ -11.8}$ & $ 134.1^{ +41.9}_{-148.8}$ & $ 15.76^{+13.56}_{ -3.50}$ & $  0.94^{ +1.10}_{ -0.30}$ & $ 0.609^{+0.125}_{-0.212}$ & $  12.3^{ +13.6}_{  -2.7}$ \\ 
          A209 & 0.2069 &  844.3 & $  29.6^{ +15.9}_{  -8.7}$ & $  10.4^{ +22.4}_{ -16.9}$ & $ 14.94^{ +7.54}_{ -4.46}$ & $  2.38^{ +1.07}_{ -0.79}$ & $ 0.428^{+0.082}_{-0.188}$ & $  16.3^{ +10.3}_{ -17.8}$ \\ 
RXJ0142 & 0.2803 &  827.5 & $   2.1^{ +24.7}_{  -2.9}$ & $  -7.1^{ +15.7}_{  -5.4}$ & $  4.68^{ +1.68}_{ -1.53}$ & $  5.87^{ +6.10}_{ -2.37}$ & $ 0.799^{+0.030}_{-0.259}$ & $ -53.6^{  +0.8}_{  -6.9}$  \\ 
          A267 & 0.2300 &  749.0 & $  -0.0^{  +7.7}_{  -8.6}$ & $   5.8^{ +16.2}_{ -12.0}$ & $  3.27^{ +0.66}_{ -0.90}$ & $ 10.04^{+11.74}_{ -2.44}$ & $ 0.769^{+0.069}_{-0.094}$ & $  26.5^{  +3.2}_{  -5.5}$  \\ 
          A291 & 0.1960 &  807.3 & $  11.2^{  +9.9}_{ -22.6}$ & $ -48.0^{ +33.4}_{ -19.6}$ & $  5.60^{ +2.86}_{ -1.52}$ & $  2.94^{ +1.75}_{ -1.09}$ & $ 0.771^{+0.062}_{-0.094}$ & $  33.9^{  +3.8}_{  -4.6}$  \\ 
          A383 & 0.1883 &  840.2 & $  -4.2^{  +2.6}_{  -3.2}$ & $  -3.3^{  +0.9}_{ -16.2}$ & $  2.01^{ +0.28}_{ -0.15}$ & $ 34.14^{ +5.55}_{ -8.92}$ & $ 0.814^{+0.067}_{-0.072}$ & $  -5.3^{  +1.6}_{  -1.6}$  \\ 
          A521 & 0.2475 &  818.3 & $   0.3^{ +19.2}_{ -62.3}$ & $ -54.6^{ +31.9}_{ -42.8}$ & $ 13.63^{ +7.76}_{ -5.35}$ & $  1.29^{ +0.70}_{ -0.61}$ & $ 0.176^{+0.206}_{-0.175}$ & $  -7.4^{ +91.3}_{ -80.1}$  \\ 
          A586 & 0.1710 &  838.7 & $  -9.0^{ +16.5}_{ -18.8}$ & $ -10.2^{ +14.6}_{  -9.7}$ & $  6.16^{ +3.49}_{ -2.52}$ & $  8.67^{ +9.37}_{ -3.72}$ & $ 0.426^{+0.212}_{-0.248}$ & $ -87.7^{ +25.8}_{ -15.2}$  \\ 
 ZwCl0740 & 0.1114 &  825.7 & $ -46.4^{ +21.9}_{ -14.4}$ & $  44.9^{ +26.0}_{ -28.7}$ & $ 18.65^{+20.79}_{ -6.74}$ & $  1.57^{ +0.82}_{ -0.62}$ & $ 0.553^{+0.117}_{-0.147}$ & $  19.3^{  +7.0}_{ -10.0}$  \\ 
ZwCl0823 & 0.2248 &  867.4 & $  45.9^{ +25.8}_{ -14.5}$ & $  26.5^{ +10.6}_{ -10.3}$ & $  9.78^{ +7.25}_{ -3.71}$ & $  2.75^{ +1.91}_{ -1.22}$ & $ 0.144^{+0.258}_{-0.144}$ & $ -47.3^{+137.2}_{ -42.4}$  \\ 
         A611 & 0.2880 &  779.7 & $ -17.5^{ +22.1}_{ -10.6}$ & $  25.9^{ +13.9}_{ -24.7}$ & $ 11.08^{ +7.91}_{ -2.70}$ & $  1.95^{ +0.94}_{ -0.97}$ & $ 0.517^{+0.127}_{-0.110}$ & $  37.1^{ +11.4}_{  -9.1}$  \\ 
          A689$^a$ & 0.2793 &  841.1 & $-184.4^{ +30.2}_{ -22.8}$ & $ 397.8^{+186.2}_{ -22.2}$ & $  9.11^{ +2.98}_{ -3.26}$ & $  0.41^{ +0.54}_{ -0.18}$ & $ 0.691^{+0.143}_{-0.087}$ & $  21.6^{  +6.2}_{  -9.2}$  \\ 
          A697 & 0.2820 &  776.9 & $   5.8^{ +10.8}_{ -16.0}$ & $ -20.9^{ +14.6}_{ -14.1}$ & $ 17.54^{ +7.27}_{ -6.16}$ & $  2.06^{ +1.18}_{ -0.72}$ & $ 0.156^{+0.170}_{-0.155}$ & $ -52.1^{+141.8}_{ -37.5}$  \\ 
          A750$^a$ & 0.1630 &  835.5 & $  10.1^{ +19.2}_{ -14.7}$ & $  28.3^{ +23.6}_{ -21.0}$ & $ 51.81^{+142.88}_{-10.60}$ & $  0.00^{ +0.01}_{ -0.00}$ & $ 0.065^{+0.140}_{-0.011}$ & $  77.2^{  +5.0}_{  -3.8}$  \\ 
       A1835 & 0.2528 &  794.8 & $  -2.0^{ +12.5}_{ -15.9}$ & $  -0.9^{ +16.7}_{  -8.4}$ & $ 11.60^{ +5.15}_{ -2.02}$ & $  3.66^{ +1.05}_{ -1.14}$ & $ 0.420^{+0.120}_{-0.180}$ & $  45.5^{ +10.1}_{ -14.7}$  \\ 
 ZwCl1454 & 0.2578 &  745.9 & $ -24.3^{  +6.8}_{ -13.2}$ & $  81.7^{ +16.7}_{ -36.3}$ & $  4.06^{ +2.88}_{ -1.39}$ & $  2.68^{ +1.81}_{ -1.24}$ & $ 0.568^{+0.106}_{-0.190}$ & $ -16.1^{ +20.6}_{ -15.7}$  \\ 
ZwCl1459$^a$ & 0.2897 &  977.9 & $ -22.3^{  +5.6}_{  -8.2}$ & $ -10.5^{ +18.6}_{ -17.5}$ & $  5.11^{ +1.23}_{ -1.62}$ & $  6.46^{ +6.60}_{ -1.55}$ & $ 0.747^{+0.090}_{-0.105}$ & $  -9.6^{  +5.6}_{  -5.3}$  \\ 
        A2219 & 0.2281 &  753.0 & $  -9.7^{ +13.0}_{ -19.8}$ & $   5.9^{  +7.3}_{  -8.4}$ & $  6.65^{ +2.69}_{ -1.70}$ & $  8.99^{ +6.33}_{ -3.11}$ & $ 0.705^{+0.043}_{-0.175}$ & $ -78.2^{  +5.5}_{  -8.0}$  \\ 
 RXJ1720$^a$ & 0.1640 &  926.5 & $ -11.7^{ +13.5}_{  -6.9}$ & $  26.2^{  +9.5}_{ -12.6}$ & $  3.37^{ +1.72}_{ -0.99}$ & $ 11.70^{+10.48}_{ -4.29}$ & $ 0.508^{+0.169}_{-0.309}$ & $  12.0^{ +57.9}_{ -10.0}$  \\ 
         A2261$^a$ & 0.2240 &  906.1 & $ -18.3^{ +11.7}_{ -10.8}$ & $  17.1^{ +10.7}_{  -6.8}$ & $ 10.45^{ +2.83}_{ -2.79}$ & $  5.16^{ +2.08}_{ -1.05}$ & $ 0.255^{+0.156}_{-0.214}$ & $  64.9^{ +31.3}_{ -27.6}$  \\ 
RXJ2129$^a$ & 0.2350 &  755.4 & $   5.0^{ +37.7}_{ -22.8}$ & $   6.6^{ +30.1}_{ -17.2}$ & $ 26.65^{+41.87}_{ -8.52}$ & $  0.00^{ +0.04}_{ -0.00}$ & $ 0.077^{+0.168}_{-0.023}$ & $  59.0^{  +6.7}_{  -7.2}$  \\ 
         A2390 & 0.2329 &  735.7 & $ -13.0^{ +16.7}_{ -15.4}$ & $ -16.5^{ +15.6}_{  -7.3}$ & $  7.27^{ +3.26}_{ -1.98}$ & $  6.39^{ +4.29}_{ -2.31}$ & $ 0.598^{+0.083}_{-0.149}$ & $ -45.0^{  +7.5}_{ -13.5}$  \\ 
         A2485 & 0.2472 &  741.4 & $ -40.6^{  +7.6}_{  -6.9}$ & $  -4.0^{ +20.6}_{  -9.4}$ & $  2.29^{ +1.33}_{ -0.34}$ & $ 15.25^{ +6.40}_{ -8.20}$ & $ 0.808^{+0.015}_{-0.210}$ & $  -8.6^{  +3.1}_{  -5.9}$  \\ 
         A2631 & 0.2780 &  816.2 & $   1.6^{ +10.5}_{  -9.6}$ & $  -3.8^{ +11.9}_{  -8.3}$ & $  4.21^{ +1.89}_{ -1.18}$ & $  9.84^{ +8.19}_{ -4.52}$ & $ 0.347^{+0.244}_{-0.344}$ & $  78.6^{ +93.1}_{ -44.5}$  \\ 
 \hline
 \end{tabular}
\flushleft{$^a$ Removed for the analysis of cluster ellipticities,
  because the elliptical NFW model does not give an acceptable 
 fit to the measurement (see text for more details).}
\end{table*}

\begin{figure*}
\begin{center}
 \includegraphics[width=0.8\hsize]{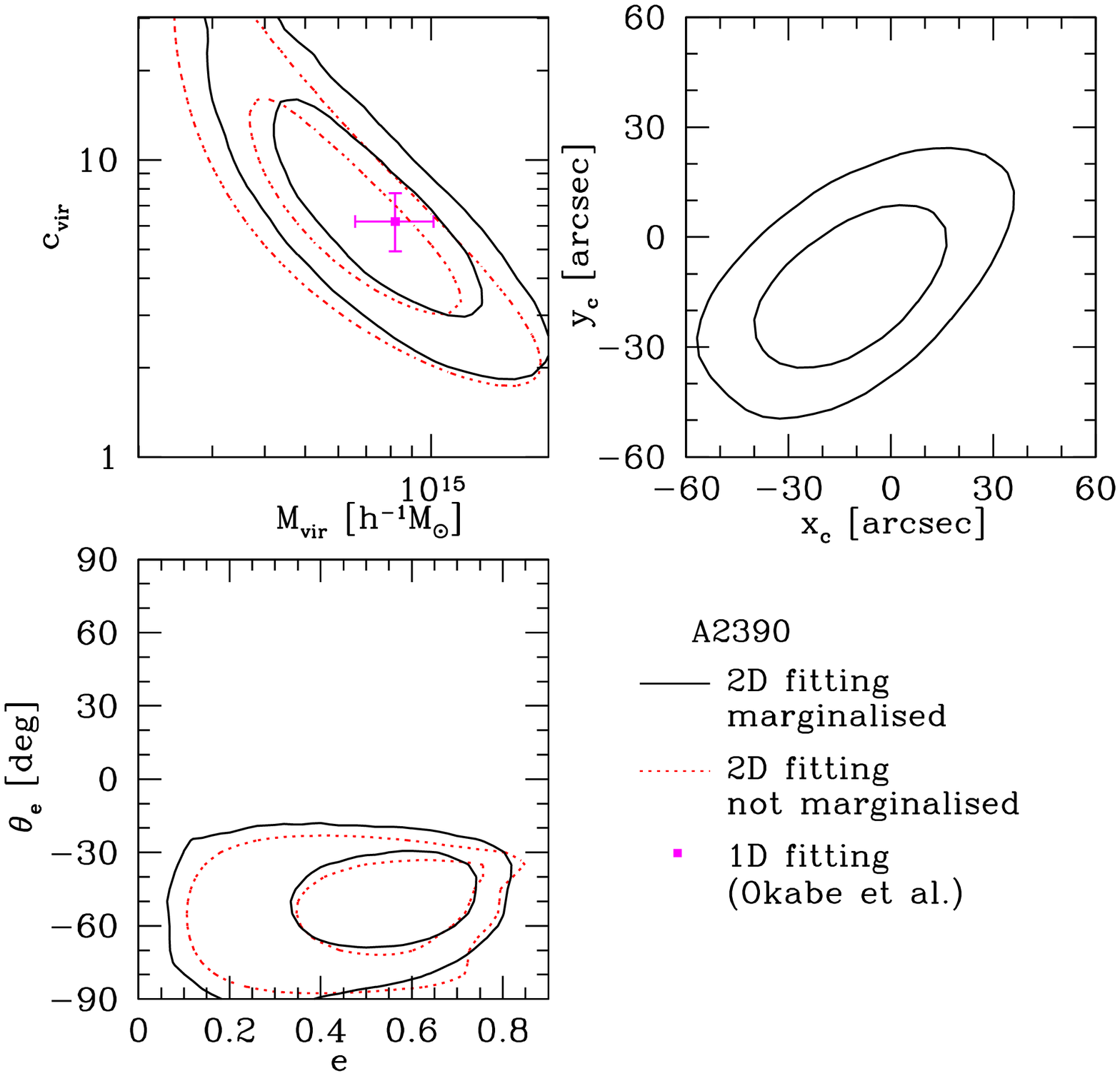}
\end{center}
\caption{
The fitting result for A2390. Solid contours show the $1\sigma$ and
$2\sigma$ error contours, marginalised over other parameters, in each
two-dimensional parameter space: the halo mass -- concentration plane
({\it upper-left}), the halo centre offsets from the BCG position
({\it upper-right}), and the halo ellipticity and its position angle
({\it lower-left}).  Dotted contours in the upper left panel show
the $1\sigma$ ($\Delta\chi^2=2.3$) and $2\sigma$ ($\Delta\chi^2=6.17$)
contours obtained assuming a spherical NFW model with the BCG position
as the cluster centre (i.e., $x_{\rm c}=y_{\rm c}=e=0$). The filled
square with error bars is the result of 1D fitting by \citet{okabe10};
the error bar appears to be small compared with the $1\sigma$ contour
of the 2D result partly because those errors are obtained from the
likelihood distribution projected onto one-dimensional space (i.e.,
$\Delta\chi^2=1$ for $1\sigma$). Dotted contours in the lower left
panel are those obtained when the mass and concentration parameters
are fixed to the best-fit values and the centre fixed to the BCG
position. The small differences between the solid and dotted contours
in these two panels suggest that the halo ellipticity and centroid
are not largely degenerate with the halo mass and concentration
parameter. 
\label{fig:cont_a2390}}
\end{figure*}

\subsection{Cluster mass model}
\label{sec:model}

We assume that the cluster mass distribution can be described by a
single halo component with its radial profile being so-called NFW
density profile \citep{navarro96,navarro97}. The spherical NFW
model is fully specified by two parameters, the halo concentration and
mass parameters. Although the density profile was obtained from
$N$-body simulation by spherically averaging the halo mass
distribution, we can construct an elliptical lens model simply by
introducing an ellipticity in the isodensity contour. Specifically, we
adopt the following mass model in our analysis:  
\begin{equation}
\kappa(x, y)=\kappa_{\rm sph}(\zeta),
\label{eq:ellnfw}
\end{equation}
\begin{equation}
\zeta^2=\frac{x'{}^2}{1-e}+(1-e)y'{}^2,
\end{equation}
\begin{equation}
x'=x\cos\theta_e+y\sin\theta_e,
\end{equation}
\begin{equation}
y'=-x\sin\theta_e+y\cos\theta_e,
\end{equation}
where $\kappa_{\rm sph}(r)$ is the radial convergence profile for the
spherical NFW profile \citep[e.g.,][]{bartelmann96}. 
Here the coordinate origin is taken as the halo centre ($x_{\rm c}$,
$y_{\rm c}$). The halo 
ellipticity $e$ is related with the major ($a$) and minor ($b$) axis
lengths of the isodensity contour as $e=1-b/a$. Throughout the paper
we adopt the coordinate system with the $x$- and $y$-axes being
aligned with West and North respectively. With this coordinate system
the position angle $\theta_e$ is measured East of North. The
lensing shear is computed by solving the two-dimensional Poisson
equation whose source term is given by the convergence $\kappa(x,y)$, 
as described in \citet{schramm90}.  We note that this elliptical model
includes a triaxial halo model which better describes haloes in $N$-body 
simulations than the spherical model \citep{jing02,kasun05,allgood06},
because the convergence map of a triaxial halo has elliptical
isodensity contours when projected along arbitrary directions
\citep{oguri03,oguri04}.  

In summary, the elliptical NFW model is specified by 6 model parameters: 
\begin{equation}
\bm{p}\equiv \left\{
M_{\rm vir}, c_{\rm vir}, e, \theta_e, x_{\rm c}, y_{\rm c}
\right\}. 
\end{equation}
Unless otherwise specified, we adopt the virial overdensity $\Delta_{\rm
 vir}\simeq 110$ (with respect to the critical density of the
universe) computed using the spherical collapse model to define
the mass $M_{\rm vir}$ and the concentration parameter $c_{\rm vir}$.

\subsection{2D weak lensing fitting}
\label{sec:2dfit}

The two-dimensional pixelised distortion field described in
\S~\ref{sec:wldata} can be compared with the two-dimensional mass model
in \S~\ref{sec:model} in order to constrain properties of halo mass
distribution including the halo ellipticity. In this paper we employ the
$\chi^2$ fitting given as 
\begin{eqnarray}
  \chi^2&=&\sum_{\alpha,\beta=1}^2\sum_{k,l=1}^{N_{\rm pixel}}
\left[g_\alpha(\bm{\theta}_k)-g^{\rm m}_\alpha(\bm{\theta}_k;\bm{p})
\right]
\left[
\bm{C}^{-1}\right]_{\alpha\beta, kl}\nonumber\\
&&\times \left[g_\beta(\bm{\theta}_l)-g^{\rm
	  m}_\beta(\bm{\theta}_l;\bm{p})
\right],
\label{eq:chi2}
\end{eqnarray}
where the indices $\alpha$ and $\beta$ run over the two components of
distortion ($\alpha,\beta=1,\,2$) and the indices $k$ and $l$ denote
the pixel position ($k,l=1,\,\ldots,\,N_{\rm pixel}$). The matrix
$\bm{C}$ denotes the error covariance matrix and $\bm{C}^{-1}$ is
the inverse matrix (see below).The best-fit model parameters are found
by minimizing the $\chi^2$ value given the distortion data.  

For the covariance matrix we include two contributions: the intrinsic
ellipticity noise, which is the primary source, and the cosmic shear
contamination arising from large-scale structures at different
redshifts, but along the same line of sight. The covariance matrix is
expressed as  
\begin{equation}
\bm{C}=\bm{C}^{\rm shape}+\bm{C}^{\rm lss}.
\label{eq:noisemt}
\end{equation}

The intrinsic ellipticity noise is expected to be uncorrelated between
different galaxies. Therefore, for the pixelised map, the shape noise 
covariance matrix has only diagonal terms: 
\begin{equation}
\left[\bm{C}^{\rm shape}\right]_{\alpha\beta,kl}=
\delta^K_{\alpha\beta}\delta^K_{kl}
\sigma^2_{g}(\bm{\theta}_k),
\end{equation}
where $\delta^K_{\alpha\beta}$ or $\delta^K_{kl}$ denotes the Kronecker
delta function, and $\sigma_g$ is given by equation~(\ref{eq:shapenoise}). 

On the other hand, the covariance due to large-scale structure is
given in \citet{dodelson04} as \citep[see also][]{hoekstra03} 
\begin{equation}
\left[\bm{C}^{\rm lss}\right]_{\alpha\beta,kl}
=\xi_{\alpha\beta}(|\bm{\theta}_k-\bm{\theta}_l|),
\end{equation}
where $\xi$ is the cosmic shear correlation functions, and we have
assumed that $\xi$ is given as a function of the length of vector
connecting the two points $\bm{\theta}_k$ and $\bm{\theta}_l$ due to the
statistical isotropy of the Universe. Specifically, the shear
correlation functions constructed from combinations of two shear
components are expresses as 
\begin{eqnarray}
  \xi_{11}(r)&=&(\cos 2\phi)^2\xi_{++}(r)+(\sin 2\phi)^2
\xi_{\times\times}(r),\\
  \xi_{22}(r)&=&(\sin 2\phi)^2\xi_{++}(r)+(\cos 2\phi)^2
\xi_{\times\times}(r),\\
  \xi_{12}(r)&=&\cos 2\phi\sin 2\phi \,\left[\xi_{++}(r)-
\xi_{\times\times}(r)\right],
\end{eqnarray}
where $r=|\bm{\theta}_k-\bm{\theta}_l|$, $\phi$ is the position angle
between the coordinate $x$-axis and the vector 
$\bm{\theta}_k-\bm{\theta}_l$, and $\xi_{++}$ and $\xi_{\times\times}$
denote the tangential and cross component shear correlation functions
\citep[e.g.,][]{bartelmann01}. For a given cosmological model, 
$\xi_{++}$ and $\xi_{\times\times}$ are computed once the nonlinear
mass power spectrum $P_\delta(k)$ and the mean source redshift are
given. We assumed the concordance $\Lambda$CDM model \citep{komatsu09}
and used the fitting formula in \citet{smith03} to compute the
nonlinear mass power spectrum. 

Note that we ignored the error contribution arising from structures
surrounding a cluster of interest. The surrounding structures cause the
so-called two-halo contribution to the shearing effect on background
galaxies \citep{sheldon07a}. However, we have checked that the two-halo
term contribution is negligible over angular scales we have
considered.

The dimension of the covariance matrix is given by twice of the number
of pixels in the distortion map. For our case, 
$2\times N_{\rm pixel}=792$. On the other hand, the total number of
parameters of an elliptical NFW model is 6, as described in
\S\ref{sec:model}. Therefore the degree of freedom of our 2D fitting 
is 786. 

In order to find the best-fit model, we need to monitor the $\chi^2$
values in 6-dimensional parameter space, which is computationally
time-consuming if using the grid based computation. 
We thus employ a Markov Chain Monte Carlo (MCMC) approach to explore
the $\chi^2$ surface, where a standard Metropolis-Hastings sampling
with the multivariate-Gaussian as a proposal distribution is adopted. 
We restrict the range of the ellipticity to $0<e<0.9$, and the
concentration as $c_{\rm vir}<40$. Beside these we do not add any
prior information. The constraints on individual parameters are
obtained by projecting the likelihood distributions (i.e., minimising
$\chi^2$) to the parameter space, including marginalisation over other
parameter uncertainties. 

\section{Results}
\label{sec:result}

\subsection{Results for individual clusters}

We study all the 25 clusters with colour information using the method
described in \S\ref{sec:fit}. Table~\ref{table:fit} summarises
best-fit parameters for each cluster, which shows several interesting
results. First, the 2D shear fitting can constrain the angular
position of halo centre reasonably well with the typical accuracy of
$\sim 20$~arcseconds ($\sim 50h^{-1}$~kpc) in radius. 
Secondly, the halo
ellipticity is detected at a high significance for many of the clusters. 
Thirdly, the ability to constrain the cluster mass and concentration
is not very degraded by including the cluster centre position and
the halo ellipticity as new fitting parameters, if we compare the
results with the 1D fitting results in \citet{okabe10}.  
Below we explore these results in more detail. 

In order to have a better understanding on parameter degeneracies, 
Figure~\ref{fig:cont_a2390} shows the projected constraint contours
in each subspace of two parameters for A2390, one of the
best-constrained clusters in our sample. We confirm that constraints
on halo mass and concentration parameter are almost unchanged even if 
the other parameters are fixed to the best-fit values or to the values
obtained from 1D fitting, i.e., assuming a spherical NFW model and the BCG
position for the halo centre. Similarly, constraints on the halo
ellipticity and its position angle is not degraded by fixing the other
parameters. These results suggest that the ellipticity is 
degenerate neither with the cluster mass/concentration nor the cluster
centre position. Therefore, by performing the full 2D fitting we gain
additional information on cluster properties, which are not obtained
from the conventional tangential shear profile,
without degrading the ability to measure the radial mass profile.

Although most clusters are fitted reasonably well, some clusters
have the large minimum $\chi^2$ value compared with the
degree-of-freedom, implying that an elliptical NFW model 
is not a good fit to the data.  For instance, the best-fit models of A750 and
RXJ2129 have very large $M_{\rm vir}$ and small $c_{\rm vir}$. The
cluster centres are not constrained very well for A115 and A689
presumably because of their multimodal mass structures \citep[see][for
  the reconstructed mass maps]{okabe10}. In addition, for RXJ1720,
A2261, and ZwCl1459, their reduced $\chi^2$ values are large  
enough, $\chi^2_{\rm min}/786\ge 1.146$, implying that the elliptical
NFW model is rejected at 3$\sigma$ level. In what follows, we
remove these 7 clusters and analyse remaining 18 clusters (except for
the analysis of the cluster centre presented in
\S\ref{sec:offset}).

One may argue that removal of these clusters from the analysis might
bias results on, e.g., cluster ellipticities which are thought to be
correlated with the complexity  of the internal mass distributions. 
However, note that the clusters excluded from the analysis do not
necessarily have complicated internal mass distributions. We find that
clusters sometimes yield poor fitting results simply because the data
quality is not good enough for our six-parameter fitting. Indeed, some
of the excluded clusters do have shallower depths or poorer seeing
sizes than others. Another reason for poor fitting includes the
projection of massive haloes along the line-of-sight, as in the case
of A2261 \citep{okabe10}. Thus we expect that the bias caused by the
exclusion is not so significant.  

\begin{figure}
\begin{center}
 \includegraphics[width=0.85\hsize]{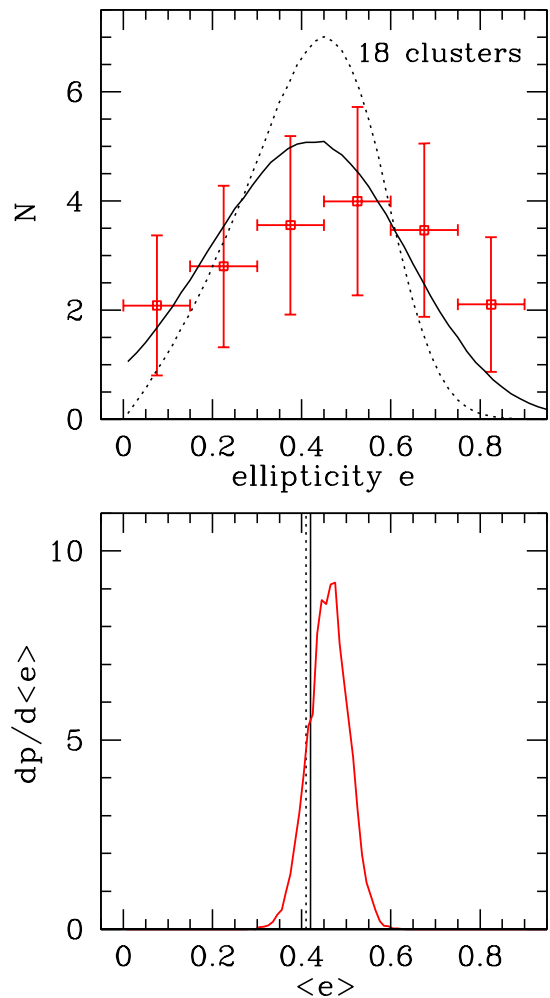}
\end{center}
\caption{{\it Upper panel:} The distribution of the halo ellipticity $e$
 for 18 clusters. Note that the ellipticity is for the projected mass
 density (see equation~[\ref{eq:ellnfw}]). The open squares with error
 bars show the observed distribution estimated from 10,000 Monte Carlo
 redistributions of the ellipticity parameters based on the posterior
 likelihood function of ellipticity for each cluster, where the error
 bars in each ellipticity bin denote the range including the 68
 percentile of 10,000 realisations. Note that different bins are not
 independent but are correlated with each other. The solid curve is the
 theoretically-expected distribution of ellipticity based on a
 triaxial halo model of \citet{jing02}, computed adopting the cluster
 redshift of $0.23$ and mass of $7\times 10^{14}h^{-1}M_\odot$ (median
 redshift and mass for our cluster sample) and convolved with the
 Gaussian with $\sigma=0.15$ which corresponds to a typical
 measurement uncertainty for our 2D shear fitting. The dotted curve
 shows the original theoretical prediction without the Gaussian
 convolution. {\it Lower panel:} The probability distribution of the
 mean ellipticity $\langle e \rangle$ for the 18 clusters. The
 vertical solid and dotted lines indicate the mean ellipticity
 expected from the triaxial halo model, with and without the Gaussian
 smoothing, respectively. 
\label{fig:ell}}
\end{figure}

\subsection{Distribution of halo ellipticities for 18 clusters}
\label{sec:ellip}

We combine the results of all the 18 clusters in order to confront 
the observed distribution of halo ellipticity parameter $e$ with the
CDM predictions. We construct the observed distribution as follows. As
we have described so far, the halo ellipticity is constrained for each
cluster, including marginalisation over other parameter
uncertainties. However, the halo ellipticity is constrained having the
projected errors. In order to properly take into account the errors,
for each cluster we randomly select the likely true value of halo
ellipticity assuming the posterior likelihood function of $p(e)\propto
\exp(-\Delta\chi^2/2)$. We generate 10,000 realisations of the
ellipticity measurement for each cluster, and combine the results of
all the clusters to estimate the mean number of clusters and its
variance in each ellipticity bin.  

\begin{figure}
\begin{center}
 \includegraphics[width=0.85\hsize]{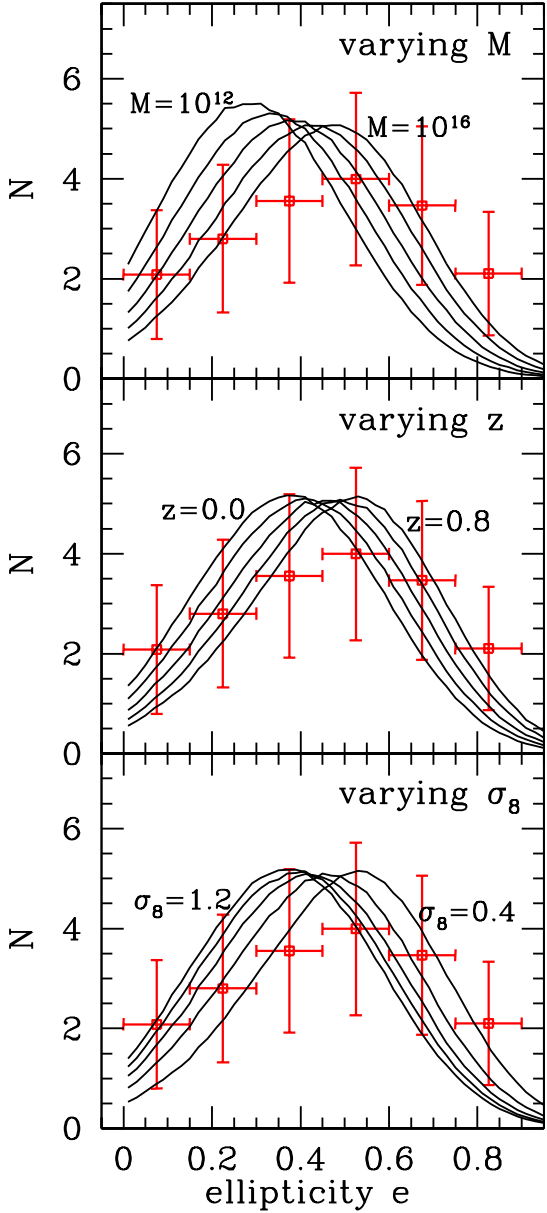}
\end{center}
\caption{The solid curves in each panel show how the
  Gaussian-convolved CDM predictions \citep[][see also
    Figure~\ref{fig:ell}]{jing02} for the halo ellipticity
  distribution change with halo mass ({\it upper}, from $M_{\rm
    vir}=10^{12}h^{-1}M_\odot$ to $10^{16}h^{-1}M_\odot$ with
  intervals of 1~dex), redshift ({\it middle}, from $z=0$ to $0.8$
  with intervals of $0.2$) and the normalisation parameter of 
  primordial density fluctuations, $\sigma_8$ ({\it lower}, from
  $\sigma_8=0.4$ to $1.2$ with intervals of $0.2$). In each panel,
  except for varying parameter, the other parameters are fixed to the
  fiducial values. The square symbols with error bars are the measured
  distribution for 18 clusters shown in Figure~\ref{fig:ell}. 
\label{fig:ell_paras}}
\end{figure}

Figure~\ref{fig:ell} shows the resulting distribution, where the error
bar in each bin denotes the range including the 68 percentile of
10,000 realisations. The distribution can be compared with the CDM
prediction of \citet{jing02}, which is derived by fitting  
dark haloes in $N$-body simulations with a triaxial NFW model. 
The CDM prediction in the Figure (dotted line) is computed as
follows. First, again note that, if projecting the 3D triaxial model
along the line-of-sight, the resulting mass distribution on the sky is
exactly the same as that given in
equation~(\ref{eq:ellnfw}). \citet{jing02} derived the probability
distribution function of triaxial halo shapes (axis ratios) as a
function of halo mass and cosmological models. Thus the theoretical
prediction for the halo ellipticity distribution of 2D mass density
can be computed by projecting the triaxial halo model along arbitrary
line-of-sight directions as described in detail in
\citet{oguri03}. It should be noted that the theoretical distribution
rests on the implicit assumption that the cluster sample is unbiased
in terms of both the shape and orientation. In the calculation, the
concordance $\Lambda$CDM model is assumed, and the mass and redshift
are fixed to the median mass and redshift of our cluster sample:
$z=0.23$ and $M_{\rm vir}=7\times 10^{14}h^{-1}M_\odot$, respectively.  

The plot shows that both observed and theoretical distributions peak
at $e\sim 0.4-0.5$, but the observed distribution is significantly
wider than the theoretical distribution. Apparently this is because of
the measurement uncertainty which broadens the distribution. Thus a
correction to the theoretical prediction is required to account for
the measurement uncertainty. The solid curve shows the the theoretical
distribution convolved with the Gaussian function with width of
$\sigma=0.15$, which corresponds to the typical uncertainty of the
ellipticity measurement for our 2D shear fitting (see
Table~\ref{table:fit}). Indeed, we find that the Gaussian-smoothed
theoretical distribution better matches the observed distribution.

While the detection of non-zero halo ellipticity may be obvious from
the distribution in Figure~\ref{fig:ell}, we can quantify how well
the elliptical model improves a fit to the 2D shear map compared to
the spherical model by monitoring the $\chi^2$ values in
equation~(\ref{eq:chi2}). The elliptical model improves the total
$\chi^2$ value for 18 clusters by $\Delta\chi^2=51$ compared with
the spherical model with $e=0$ fitted to the same 2D data, thereby
representing the detection of an ellipticity at $7\sigma$ confidence
level.  

In the lower panel of Figure~\ref{fig:ell} the mean halo ellipticity, 
$\langle e \rangle$, for our sample of 18 clusters is compared with
the theoretical prediction, where the width of the mean ellipticity
distribution reflects the scatter among 10,000 realisations. The
observed distribution has the mean ellipticity of 
$\langle e \rangle=0.46\pm 0.04$ ($1\sigma$) which is in excellent
agreement with the triaxial model prediction, $\langle e \rangle=0.42$
(0.41) with (without) the Gaussian smoothing. 

The halo ellipticity depends on mass and redshift of clusters as well as
on cosmological models. According to the CDM hierarchical structure
formation scenario \citep{jing02}, dynamically young haloes tend to
have a more elongated shape at a given observed redshift.  In other
words, more massive haloes that have just recently formed tend to have
a larger halo ellipticity. Figure~\ref{fig:ell_paras} shows how the
theoretically expected distribution of halo ellipticity depends on
redshift and mass of haloes and one of cosmological parameters,
$\sigma_8$, the normalisation of primordial density fluctuations. 
While the current measurement is not enough to discriminate the model
differences due to a limited sample size, the Figure illustrates how
measurements of halo ellipticities can potentially 
test the structure formation model.

\subsection{Offset between lensing centre and BCG}
\label{sec:offset}

In this section, we compare the positions of cluster centres inferred
from weak lensing and from the brightest cluster galaxy(ies). Weak
lensing provides a unique method to determine the centre position of
dark matter distribution, and therefore is quite complementary to the
optical (BCGs) and X-ray based methods. It should be also noted that a
possible uncertainty in the centre position determination is currently
one of the most important systematic sources in the stacked lensing
analysis, cluster-background galaxy cross-correlation measurement
\citep{johnston07,mandelbaum10}.  
 
Our basic result, summarised in Table~\ref{table:fit}, is that the
mass centres tend to be consistent with the locations of the
BCGs. We find that mass centres are consistent with the BCG within
$2\sigma$ level for most of the clusters. However, a possible
significant deviation between the lensing and BCG centre positions is
apparent for some of the clusters.

In the following we explore a possible signature of the large offset
between BCG position and lensing centre in more detail. While we have
worked mainly on a ``clean'' subsample of 18 clusters all of which are
well fitted by an elliptical NFW model, here we analyse the full 25
clusters (all of which have colour information) in order to achieve a
more unbiased study on the offset issue.  

\begin{figure}
\begin{center}
 \includegraphics[width=0.9\hsize]{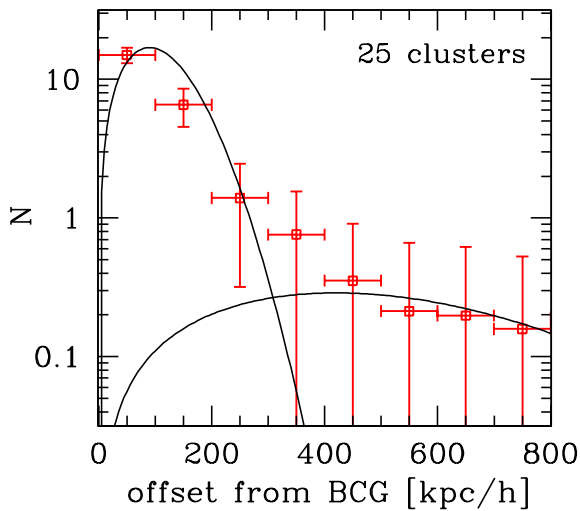}
\end{center}
\caption{The distribution of the physical distance between the halo
  centre and the BCG position from the full sample of 25 clusters,
  obtained from 10,000 Monte Carlo redistributions computed as done in
  Figure~\ref{fig:ell}. The solid curves show the two 2D
  Gaussian distributions with widths of $\sigma=90h^{-1}$~kpc and
  $420h^{-1}$~kpc, respectively, where the Gaussian is given as 
  $p(r)\propto r\exp[-r^2/(2\sigma^2)]$. The narrower Gaussian is
  obtained by fitting to the measured distribution. On the other hand,
  the broader Gaussian models a contribution with large offsets,
  motivated by the studies of \citet{johnston07}. The width is kept
 fixed to $420h^{-1}$~kpc according to \citet[][see also text
 for the details]{johnston07}, but its normalisation, treated as a
 free parameter, is estimated so as to best match the measured
 distribution.  
\label{fig:dis}}
\end{figure}

Figure~\ref{fig:dis} shows the lensing constrained distribution of the
offset amount of the cluster centre from the BCG position for the 25
clusters. The distribution is computed from 10,000 Monte Carlo
redistributions as done in Figure~\ref{fig:ell}.  First, the observed
distribution generally shows that the BCG position is close to the
cluster centre constrained from lensing. However, the distribution
also indicates a tail at large offsets, although the significance is
weak.  To be more quantitative, the observed distribution is compared
with a combination of the two 2D Gaussian distributions (sold curves)
each of which is modeled as $p(r)\propto r\exp[-r^2/(2\sigma^2)]$ ($r$
is the distance between the halo centre and BCG position). This form
is motivated by the study in \citet{johnston07}, where the
cross-correlation weak lensing measurement between clusters and
background galaxy images was studied in great detail using the SDSS
data. They used the BCG position as the halo centre to carry out the
stacked lensing analysis, and took into account the miscentring
effect of the possible BCG offset by using the 2D Gaussian form with
$\sigma=420h^{-1}$kpc. This form was chosen based on their mock galaxy 
catalogues, from which they argued that clusters may be categorised
into two populations, BCG-centred and BCG-offsetted clusters, and that
the projected radial offset for the latter population can be described
by the 2D Gaussian distribution. The two solid curves in
Figure~\ref{fig:dis} shows the best-fit two 2D Gaussian distributions
to the observed distribution, where the width of the broader Gaussian
function is kept fixed to $\sigma=420h^{-1}$ motivated by the 
study of \citet{johnston07}.  One can find that the
best-fit curve fairly well matches the observed distribution. 
In addition, the relative strength of the two 2D Gaussian distributions
implies a $10\%$ fraction of BCG-offsetted clusters, which is again
consistent with the assumption used in \citet{johnston07}.

While the observed distribution is broadly consistent with the
theoretical expectation, it is not clear whether the BCG offset is
detected significantly, given the relatively large measurement
uncertainties. In fact, the typical error on the mass centroid
measurement is  $\sigma\sim 50h^{-1}$~kpc, and there are a few
clusters in the sample that have quite large errors, $\sigma >
100h^{-1}$~kpc, which are nonnegligible compared to the widths of the 
2D Gaussian distributions discussed above. Thus, for more quantitative
discussions on the distribution of BCG offset, we need
a larger sample of weak lensing clusters.

We also compare the offsets between BCGs and mass centroids from weak
lensing with those between BCGs and X-ray centres derived by
\citet{sanderson09} using {\it Chandra} X-ray images. We do not find
any significant correlation between these offsets. One of the reasons
is that offsets between BCGs and X-ray centres for our cluster sample
are typically much smaller than the accuracy of mass centroid
determinations from weak lensing. We leave more detailed comparisons
with X-ray data in future work.

\begin{table}
 \caption{The shapes and orientations of member galaxies and BCGs. The
   errors are the marginalised $1\sigma$ errors (see text for the details).
\label{table:gals}}   
 \begin{tabular}{@{}ccccc}
  \hline
   Name
   & $e_{\rm mem}$
   & $\theta_{\rm mem}$ 
   & $e_{\rm BCG}$ 
   & $\theta_{\rm BCG}$  \\
   &
   & (deg)
   & 
   & (deg) \\
 \hline
     A68 & $ 0.585^{+0.108}_{-0.168}$ & $ -59.5^{  +9.4}_{ -10.4}$ & $0.326$ & $-58.7$ \\ 
    A115 & $ 0.298^{+0.132}_{-0.206}$ & $  17.2^{ +27.0}_{ -10.6}$ & $0.268$ & $-30.9$ \\ 
    A209 & $ 0.495^{+0.040}_{-0.060}$ & $ -34.8^{  +3.9}_{  -3.6}$ & $0.539$ & $-54.8$ \\ 
RXJ0142  & $ 0.316^{+0.132}_{-0.184}$ & $ -46.7^{ +11.5}_{ -16.1}$ & $0.341$ & $-68.4$ \\ 
    A267 & $ 0.500^{+0.125}_{-0.082}$ & $   8.7^{  +7.2}_{  -6.1}$ & $0.507$ & $ 19.1$ \\ 
    A291 & $ 0.313^{+0.133}_{-0.172}$ & $ -51.3^{ +17.6}_{ -13.8}$ & $0.121$ & $-63.7$ \\ 
    A383 & $ 0.231^{+0.093}_{-0.084}$ & $  52.3^{ +16.1}_{ -12.5}$ & $0.085$ & $-23.2$ \\ 
    A521 & $ 0.308^{+0.076}_{-0.069}$ & $ -40.7^{  +8.9}_{  -9.1}$ & $0.262$ & $-44.1$ \\ 
    A586 & $ 0.337^{+0.142}_{-0.130}$ & $ -78.2^{ +12.7}_{ -10.1}$ & $0.157$ & $ 54.6$ \\ 
ZwCl0740 & $ 0.718^{+0.163}_{-0.152}$ & $ -64.4^{  +8.9}_{  -5.4}$ & $0.233$ & $-42.3$ \\ 
ZwCl0823 & $ 0.342^{+0.207}_{-0.248}$ & $ -25.9^{ +32.7}_{ -14.7}$ & $0.303$ & $-23.7$ \\ 
    A611 & $ 0.299^{+0.087}_{-0.105}$ & $  56.9^{ +13.1}_{ -15.2}$ & $0.220$ & $ 36.8$ \\ 
    A689 & $ 0.900^{+0.000}_{-0.013}$ & $   8.9^{  +3.0}_{  -1.5}$ & $0.471$ & $ 38.9$ \\ 
    A697 & $ 0.499^{+0.072}_{-0.086}$ & $ -38.8^{  +3.6}_{  -4.1}$ & $0.278$ & $-14.0$ \\ 
    A750 & $ 0.229^{+0.092}_{-0.064}$ & $ -81.5^{ +12.4}_{  -8.5}$ & $0.364$ & $-41.2$ \\ 
   A1835 & $ 0.562^{+0.080}_{-0.076}$ & $ -24.4^{  +4.0}_{  -5.3}$ & $0.355$ & $-40.6$ \\ 
ZwCl1454 & $ 0.518^{+0.110}_{-0.124}$ & $  12.0^{ +11.8}_{  -9.4}$ & $0.302$ & $ 35.4$ \\ 
ZwCl1459 & $ 0.209^{+0.171}_{-0.175}$ & $  43.4^{ +30.7}_{ -20.8}$ & $0.360$ & $-83.8$ \\ 
   A2219 & $ 0.340^{+0.063}_{-0.053}$ & $ -67.8^{  +7.5}_{  -5.6}$ & $0.301$ & $-58.4$ \\ 
RXJ1720  & $ 0.473^{+0.083}_{-0.147}$ & $  19.7^{  +8.8}_{ -10.4}$ & $0.189$ & $ 29.5$ \\ 
   A2261 & $ 0.407^{+0.073}_{-0.066}$ & $  48.9^{  +6.9}_{  -3.8}$ & $0.185$ & $ -1.1$ \\ 
RXJ2129  & $ 0.352^{+0.064}_{-0.075}$ & $  67.3^{  +8.7}_{  -8.1}$ & $0.443$ & $ 68.1$ \\ 
   A2390 & $ 0.625^{+0.043}_{-0.041}$ & $ -56.0^{  +3.8}_{  -2.6}$ & $0.232$ & $-50.2$ \\ 
   A2485 & $ 0.186^{+0.146}_{-0.185}$ & $ -36.1^{+124.6}_{ -51.0}$ & $0.263$ & $ 64.9$ \\ 
   A2631 & $ 0.562^{+0.078}_{-0.096}$ & $  61.6^{  +4.7}_{  -4.0}$ & $0.316$ & $ 81.1$ \\ 
 \hline
 \end{tabular}
\end{table}

\subsection{Comparison with member galaxy distributions}
\label{sec:mem}

\begin{figure*}
\begin{center}
 \includegraphics[width=0.8\hsize]{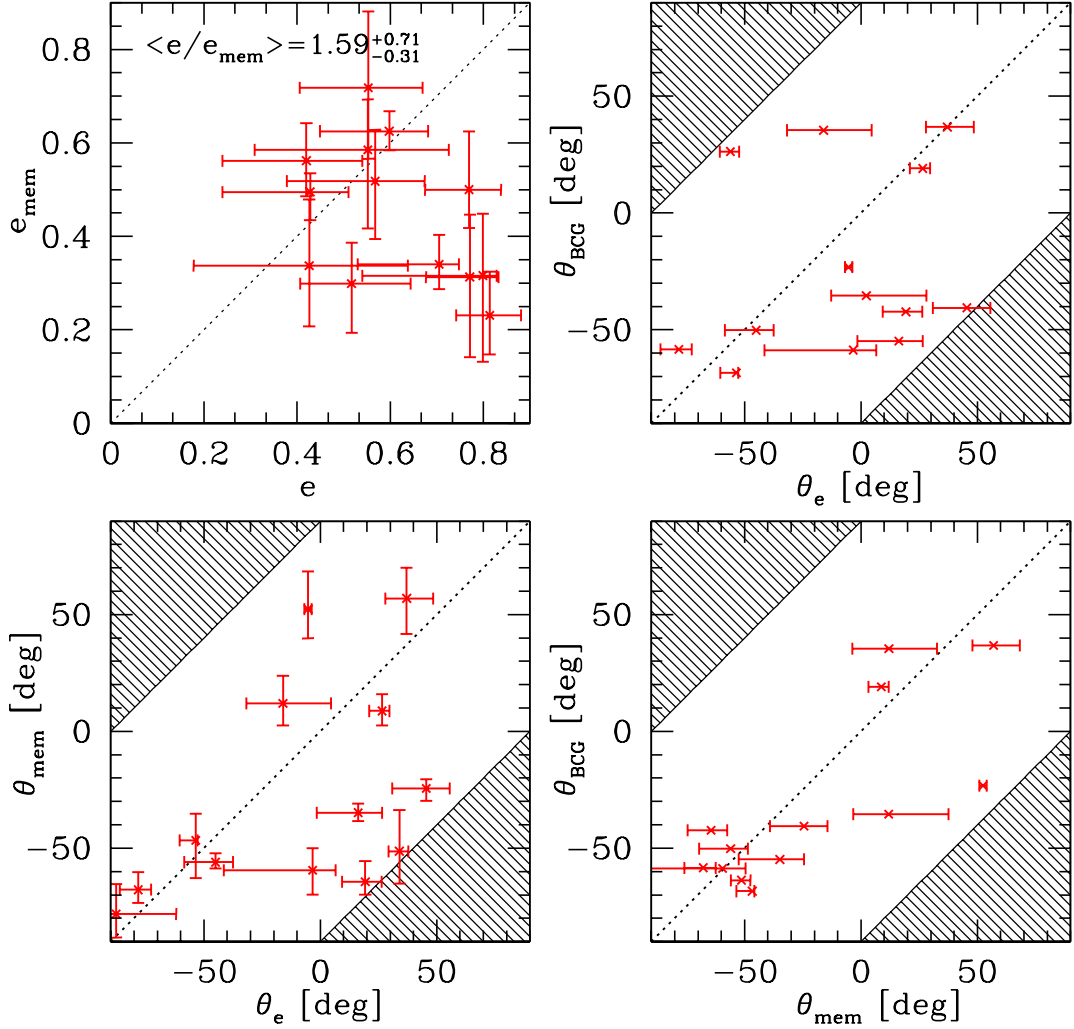}
\end{center}
\caption{Comparisons between the mass and member galaxy distributions.
  {\it Upper-left panel:} The comparison of the mass ellipticity $e$
  with the ellipticity of the member galaxy distribution, 
  $e_{\rm mem}$, for a sample of 13 clusters whose position angles are
  well constrained. The dotted line indicates $e=e_{\rm mem}$. The
  ellipticity ratio among the cluster sample is 
  $\langle e/e_{\rm mem}\rangle=1.59^{+0.71}_{-0.31}$, showing no
  strong correlation.  
  {\it Lower-left panel:} The position angles of the mass distribution
  from 2D shear fitting, $\theta_e$, are compared with the position
  angles of member galaxy distributions, $\theta_{\rm mem}$. The
  dotted line shows the case of the perfect alignment,
  $\theta_e=\theta_{\rm mem}$. The shaded corners have misalignments
  greater than $90^\circ$ and therefore are excluded (for some
  clusters both the position angles are shifted by 90~degree to fit
  in the proper region).
 {\it Upper-right panel:} Comparison between $\theta_e$ and the BCG 
 position angle $\theta_{\rm BCG}$.
 {\it Lower-right panel:} Comparison between $\theta_{\rm mem}$ and
 $\theta_{\rm BCG}$. 
\label{fig:paell_mem}}
\end{figure*}

It is often assumed that the distribution of cluster member galaxies
follows the mass distribution \citep[e.g.,][]{evans09}. 
We test this assumption based on our 2D shear fitting. 

First we defined member galaxies as galaxies sitting around the
cluster red-sequence locus in the colour-magnitude diagram with their
red-band magnitudes brighter than 22~mag \citep[see also][]{okabe10}. 
We fit the number distribution of member galaxies by assuming an
elliptical power-law profile. In fitting we fix the centre to the
best-fit centre of the mass distribution from 2D shear fitting. Since
the member galaxy catalogue is expected to be more or less
contaminated by field galaxies, we add the constant density component
in addition to the primary power-law component. Thus the parameters of
our fitting is the slope and normalisation of the power-law profile,
the ellipticity and position angle, and the constant density
component. The fitting is performed in the same $20'\times20'$ region
as in the shear fitting for each cluster. The result is summarised in
Table~\ref{table:gals}.  The table also lists the shape and
orientations of BCG for each cluster, which is obtained from
the SExtractor output \citep{bertin96}. The measurement errors for BCG
shapes are negligibly small, although the orientations of the BCGs may
slightly depend on the radius where it is determined.

Figure~\ref{fig:paell_mem} compares the 2D shapes of mass and member
galaxy distributions. Here we restrict our analysis to a subsample of
13 clusters whose position angles are constrained reasonably well 
in our fitting. First, while ellipticities mass and member galaxies
are consistent with each other for roughly half of the clusters,
overall we do not see any strong correlation between the
ellipticities. Although we find that the mass ellipticity tend to be
larger than the ellipticity of member galaxy distributions, this may
be ascribed to the selection effect: We only consider clusters with
well-determined position angles from shear fitting, which
preferentially selects clusters with large mass ellipticities, as is
clear from  Figure~\ref{fig:paell_mem}.  Is is also possible that our
ellipticity measurement of the member galaxy distribution is biased
because of the inclusion of photometrically selected red-sequence
galaxies only and the contamination by field galaxies.

Although we see some correlation between position angles of mass and
member galaxies distributions, the alignment is not very strong. In some
case the position angles are similar between the mass and member
galaxy distributions, but there are cases that the orientations of the
member galaxy distributions are significantly different from those of
the mass distributions. The results are qualitatively similar even if
we adopt orientations of BCGs. On the other hand, the distributions of
member galaxies and the orientations of BCGs are aligned better
with each other.  

To understand the origin of the misalignment, we visually inspect mass
maps of clusters which show large misalignments, and find that many of
them appear to possess significant substructures (e.g., A68, A209,
A291, A383). However, it is not obvious whether these substructures
cause the misalignment directly, or clusters with significant
substructures are dynamically young and have complicated internal mass
distributions and larger fractions of blue member galaxies which might
cause the misalignments. Indeed, our simple simulations in
Appendix~\ref{sec:app} indicate that massive subpeaks do not have
significant impact on the measurements of 2D cluster shapes unless
their masses are approximately more than 20\% of the main  halo. In
either case, our results imply that the non-sphericity of halo cannot
be reliably explored using the galaxy distribution or even using the
stacked two-dimensional lensing technique which assumes the alignment
of galaxy and dark matter distribution \citep{evans09}.   

\subsection{Cluster ellipticities from other techniques}

While our weak lensing technique provides a direct detection of
the cluster mass ellipticity over the entire cluster region,
signatures of cluster non-sphericity have been obtained using other
techniques as well. Here we review some of the previous results
in comparison with our results. 

One straightforward way to study the shape of clusters is to see the
distribution of member galaxies \citep{rhee91,detheije95,adami98,strazzullo05}.
Indeed ellipticities of $\sim 0.2-0.5$ have been detected for a number
of clusters. While member galaxies (stars) are collisionless and
therefore their spatial distribution is naively expected to follow the
underlying dark matter distribution, the complicated galaxy formation
physics makes it rather difficult to predict theoretically how these
distributions are in fact connected. On the other hand, measurements
of cluster shapes from X-ray 
\citep{mcmillian89,mohr95,boute95,kolokotronis01,flores07} may be
easier to interpret, because one can connect the distribution of X-ray
emitting hot gas with the underlying dark matter distribution under
the assumption of the hydrostatic equilibrium
\citep[e.g.,][]{lee03,wang04}, although the assumption sometimes does
not hold -- an extreme example is the bullet cluster \citep{clowe06}.
The typical observed ellipticities of X-ray surface brightnesses of
$e\sim 0.2$ are smaller than our measurement from weak lensing, which
is in fact expected because under the hydrostatic equilibrium
assumption the isodensity surface of the intra-cluster gas follows the
iso-potential surface, which is generally rounder than the dark matter
distribution \citep{buote92}. Recently, \citet{kawahara10} derived the
axis ratio distribution of X-ray clusters using XMM-Newton data.
It was claimed that, by taking into account the apparent rounder shape
in the X-ray image, the distribution agrees well with the theoretical
prediction based on the triaxial dark matter halo and hydrostatic
equilibrium, which therefore seems consistent with our finding.  

The dark matter distribution can also be studied by strong
gravitational lensing. While the necessity of large ellipticities in
some individual cluster strong lensing mass models has long been recognised 
\citep[e.g.,][]{miralda95}, recently \citet{richard10} presented a 
strong lens analysis of 20 LoCuSS clusters, and derived the mean
ellipticity of $\langle e \rangle \sim 0.4$, in broad agreement with
our result. One possible caveat is that the parameter constraints are
highly degenerated due to the nature of complex nonlinear fitting in
the strong lensing region. In addition, the ellipticity constraint
from strong lensing is on much inner region ($\la 100$~kpc in radius)
than our weak lensing measurements, where the mass distribution may be
significantly affected by various baryonic effects such as radiative
cooling and the dynamical interaction between dark matter and central
galaxies. Nevertheless, strong lensing constraints are of great
interest because of a natural prediction of the CDM model that the halo
ellipticity and position angle slightly changes with cluster centric
radii due to the collision-less nature of dark matter \citep{jing02}. 
Therefore the combination of strong lensing and weak lensing
constraints may allow a more quantitative test on these scenarios. 
Indeed  we are planning to conduct combined strong and weak lensing
analysis for our LoCuSS cluster sample, which will be presented
elsewhere (G. Smith et al., in preparation). 

\subsection{Ellipticity of galaxy-scale haloes}

Both the surface brightness distribution \citep[e.g.,][]{sheth03,choi07} 
and strong lens mass modeling \citep[e.g.,][]{keeton98,koopmans06}  
indicate that the mean ellipticity of $\langle e \rangle \sim 0.2-0.3$
near the core of early-type galaxies. The standard hierarchical
structure formation scenario suggests that more massive haloes are
more dynamically young and therefore are more elongated
\citep[e.g.,][]{jing02}. Thus the smaller ellipticities of early-type
galaxies than massive clusters are qualitatively in good agreement
with the theoretical expectation. 

However, the central mass distribution of early-type galaxies is more 
likely affected by radiative cooling and star formation of baryons
\citep{kazantzidis04,tissera10}. Constraints on the shape of
galaxy-scale haloes at outer regions, measured by stacked weak lensing
technique, are somewhat controversial 
\citep{hoekstra04,mandelbaum06a,parker07}.

\subsection{Comparison with 1D tangential shear fitting}

\begin{figure}
\begin{center}
 \includegraphics[width=0.9\hsize]{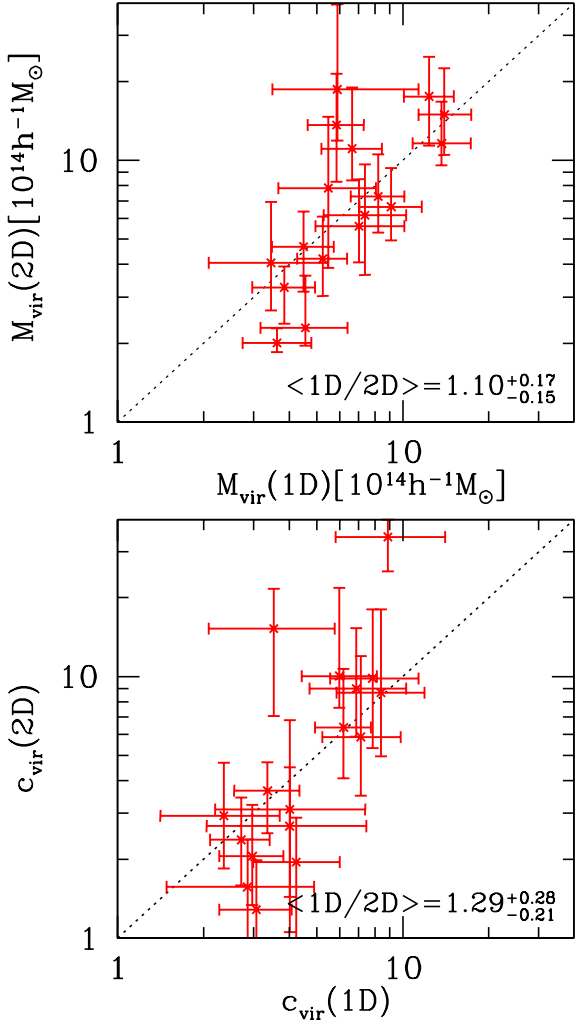}
\end{center}
\caption{The upper panel compares the best-fit mass parameters 
($M_{\rm vir}$) from the 1D shear fitting \citep{okabe10} and the 2D
  fitting in this paper, while the lower panel show the concentration
  parameters ($c_{\rm vir}$). The mean ratios are 
  $\langle M_{\rm vir}({\rm 1D})/ M_{\rm
    vir}({\rm2D})\rangle=1.10^{+0.17}_{-0.15}$ and $\langle c_{\rm 
    vir}({\rm 1D})/c_{\rm vir}({\rm2D})\rangle=1.29^{+0.28}_{-0.21}$,
 respectively.
\label{fig:mc_1d2d}}
\end{figure}

The conventional way to extract cluster mass and concentration from
weak lensing data is to fit a 1D (i.e., spherical) mass model to the
1D tangential shear profile \citep[e.g.,][]{okabe10}. We can therefore
now compare the results of our 2D model fits to the 2D shear data with
those of 1D models obtained by \citet{okabe10}. For instance, if an
elliptical halo model is a better description of real clusters than a
spherical one and if the lensing data has the ability to discriminate
these two models, the 2D shear fitting method can bring a more
reliable, unbiased estimate on cluster parameters, which is
particularly important for cluster cosmology experiments aimed at
constraining dark energy. In the following we focus on the subsample
of 17 clusters, the ``good'' 18 clusters but excluding the cluster
ZwCl0823 because  ZwCl0823 does not allow a good 1D shear fitting due
to the complex, non-spherical mass distribution \citep{okabe10}.  

Figure~\ref{fig:mc_1d2d} compares the best-fit mass/concentration
parameters obtained from the 1D and 2D fitting methods. The 1D and 2D
results agree reasonably well. In particular, the virial masses
$M_{\rm vir}$ from 1D and 2D fitting are well consistent with each
other within $1\sigma$ statistical errors. With the current sample
size we cannot conclude that there is a systematic bias in the
different cluster mass estimates. The possible deviation, $\langle
M({\rm 1D})/M(\rm {2D}) \rangle=1.10^{+0.17}_{-0.15}$, averaged
over the 18 clusters is within the $1\sigma$ scatter. 

On the other hand, the concentration parameters are more scattered, 
and the average ratio of $\langle c_{\rm vir}({\rm 1D})/c_{\rm vir}
({\rm2D})\rangle=1.29^{+0.28}_{-0.21}$ deviate from unity with $1\sigma$
confidence level. This is probably related with the fact that the
determination of concentration parameter is more difficult and is
easily affected by various uncertainties such as the fitting region
and cluster centres. In fact, the fitting regions are slightly
different between \citet{okabe10} and our 2D analysis, which can cause
such slight discrepancy.   

We note that lensing determinations of $M_{\rm vir}$ and $c_{\rm vir}$
are sometimes highly degenerate. Thus slightly different fitting method
can result in large different in $M_{\rm vir}$ and $c_{\rm}$ just
because the best-fit value moves along this degeneracy
direction. Indeed, we compare 1D and 2D results for individual
clusters, and find that cluster with $M_{\rm vir}({\rm 1D})>M_{\rm
  vir}({\rm 2D})$ tend to have $c_{\rm vir}({\rm 1D})<c_{\rm
  vir}({\rm 2D})$, being consistent with the above picture, which also
explains the relatively large scatter between 1D and 2D results seen in
Figure~\ref{fig:mc_1d2d}.

In Appendix~\ref{sec:m500} we also show the 2D fitting results for
$M_{\rm 500}$ and $c_{500}$ defined by the overdensity $\Delta=500$,
which is more relevant when comparing the lensing results with the X-ray
derived mass estimates \citep[e.g.][]{vikhlinin09}. 

\subsection{Effect of large-scale structure}

\begin{figure}
\begin{center}
 \includegraphics[width=0.9\hsize]{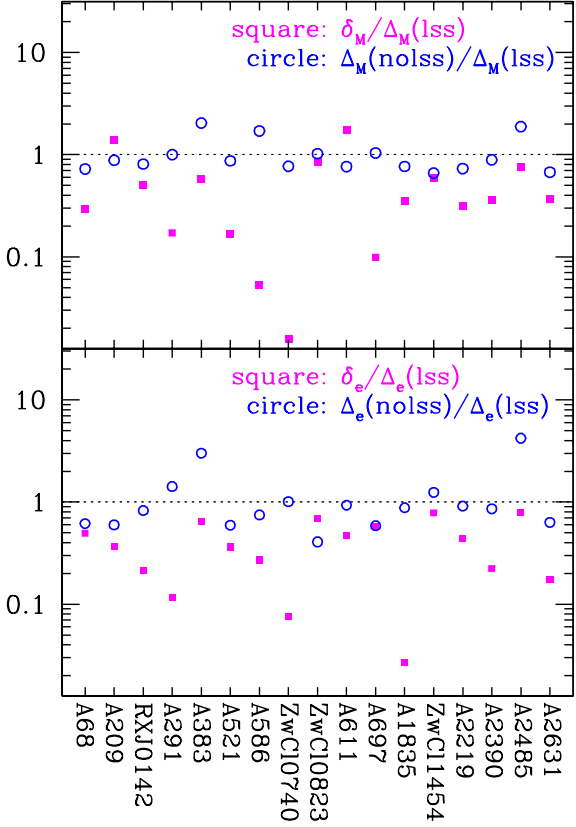}
\end{center}
\caption{
  Shown is how the cosmic shear contamination ($\bm{C}^{\rm lss}$)
  in the error covariance matrix in the 2D shear fitting (see
  equation~[\ref{eq:noisemt}]) affects estimations of 
  the mass ({\it upper panel}) and the halo ellipticity ({\it lower}). 
  We quantify this effect by the following two parameters; 
  the bias $\delta_p\equiv \left|p({\rm lss})-p({\rm
      nolss})\right|/\sigma_p({\rm lss})$, where $p$ and $\sigma_p$
  are the best-fit parameters (mass or ellipticity) and the marginalised
  $1\sigma$ errors computed  with and without the contamination 
  $\bm{C}^{\rm lss}$, and the fractional errors,
  $\Delta_p\equiv \sigma_p/p$, for both  the results with and without 
  $\bm{C}^{\rm lss}$. The ratio $\delta_p/\Delta_p({\rm lss})$, shown by 
  the square symbols, quantifies how much the best-fit parameters are
  shifted compared with the statistical errors. The open circles show
  $\Delta_p({\rm nolss})/\Delta_p({\rm lss})$ which indicate the
  change of measurement errors due to $\bm{C}^{\rm lss}$. 
\label{fig:lss}}
\end{figure}

Another new result of this paper is that we properly took into account 
the effect of cosmic shear contamination due to large-scale structure
on the estimation of cluster parameters on individual cluster basis 
(see around equation~[\ref{eq:noisemt}] for the details). While
\citet{hoekstra03} theoretically estimated that the cosmic shear
contributes to the error covariance matrix \citep[see
also][]{dodelson04}, the effect has been ignored in most of previous
cluster lensing studies, which suggests that previous work may have
underestimated errors on mass estimates. 

Figure~\ref{fig:lss} studies a bias in the best-fit parameters as well
as a change of the parameter estimation accuracy when the cosmic shear
contamination in the covariance matrix ($\bm{C}^{\rm lss}$ in
equation~[\ref{eq:noisemt}]) is included or ignored.   
Here we consider 17 clusters, the fiducial 18 clusters minus A267,
because A267 is not fitted very well without the $\bm{C}^{\rm lss}$
term in the covariance matrix. First, the Figure shows that the
parameter accuracies are little changed by including $\bm{C}^{\rm
  lss}$, while a few clusters show a large degradation. Thus the
degradation amount is not as large as that claimed in 
\citet{hoekstra03}, presumably because the number density of our 
background galaxy sample (``red+blue'' galaxies) is much smaller than
that assumed in \citet{hoekstra03} and therefore the shot noise is a
more dominant source of the measurement errors than the cosmic shear
contamination. The best-fit parameters are not largely biased
by including $\bm{C}^{\rm lss}$, but are in general consistent with
each other within statistical uncertainties. 

\section{Discussions on Systematic Errors}
\label{sec:discussion}

In this section we discuss possible systematic errors inherent in our weak
lensing measurements. 

\subsection{Constraining cluster region for the halo ellipticity}
\label{sec:radi}

\begin{figure}
\begin{center}
 \includegraphics[width=0.9\hsize]{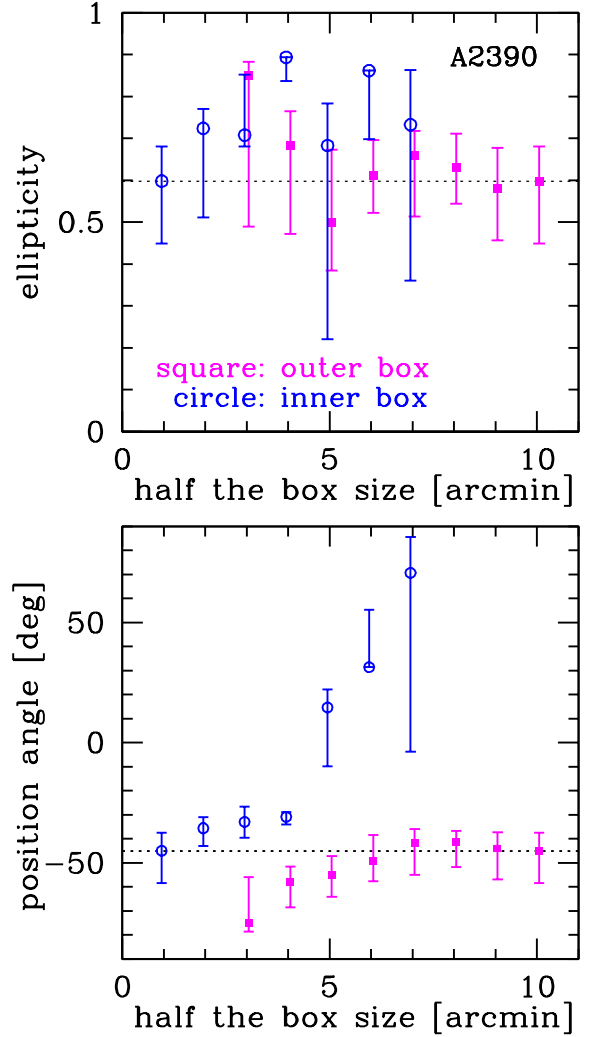}
\end{center}
\caption{The best-fit halo ellipticities ({\it upper}) and position
  angles ({\it lower}) with their marginalised $1\sigma$ errors for
  different fitting regions, for the case of A2390. We change the
  outer box size while fixing the inner box size ({\it square
    symbols}) or vice versa ({\it circles}). Note that shear fitting is
  performed in the range inside the outer box and outside the inner
  box, around the coordinate origin (taken to be the BCG position).
  The ellipticities are plotted as a function of half the box size,
  which very roughly corresponds to the radius from the cluster centre. 
\label{fig:range}}
\end{figure}

Our analysis has so far used the square region of $20'\times 20'$
centred at the BCG position, but excluding the inner region of
$2'\times 2'$ (see \S~\ref{sec:wldata}). In order to obtain a better
understanding of  the origin of the constraints on the ellipticity, we
check the dependence of the result on the cluster fitting
region. Specifically, we change the outer and inner boundary box sizes
to see how the constraint on halo ellipticity
changes. Figure~\ref{fig:range} shows  the result for the cluster
A2390. When  the outer box size is changed with fixed inner boundary, 
the best-fit ellipticity and its error are little changed as long as
the region with $\la 5'$ is included. On the other hand, 
with increasing the inner box size, but keeping the outer boundary,  the
resulting ellipticity deviates from the original best-fit value at
$\ga 3'$. The result indicates that the most important radius for
determining cluster ellipticities is $3' \la r \la 5'$ ($\sim
400-800h^{-1}$~kpc) where the weak
lensing signals tend to have maximum signal-to-noise ratios given the
cluster lensing signal and the shot noise due to the finite number
density of background galaxies in Subaru data \citep{okabe10}. 
This trend is generally true for all the ``good'' 18 clusters. 

The Figure indicates that the best-fit position angle can shift
by changing the fitting region. This may be an indicative of the twist
of isodensity contours seen in $N$-body simulations \citep{jing02},
although non-local nature of weak lensing shear makes the
interpretation somewhat difficult.

For more definiteness, we check the stability of our result on the
cluster ellipticity adopting two different boundary sizes. One is the
smaller  outer boundary box size of $10'\times10'$, and the other is
the larger inner box size of $4'\times4'$. We repeat the analysis for
these two cases, and obtain the mean ellipticities of $\langle e
\rangle=0.47\pm 0.04$ and $0.50\pm0.05$, respectively,  which are
quite consistent with the main result presented in
\S\ref{sec:ellip}. This indicates that our results are not dependent
on the specific choice of the fitting boundaries very much. 

\subsection{Dilution effect}

If the background galaxy sample for weak lensing analysis contains
a significant fraction of cluster member galaxies, it dilutes
weak lensing signals \citep{broadhurst05,medezinski07}. This dilution
effect becomes more pronounced near the cluster centre where the
number density of cluster member galaxies is higher. Since weak
lensing shear is non-local, it decreases very slowly with increasing
radius, leading to much smaller dilution effect at outer radii. 
To check the impact of the dilution effect, we repeat the same 2D
shear fitting for the ``faint'' galaxy sample (galaxies selected only by
the magnitude cut, without any selection from the colour) instead of
``red+blue'' galaxy sample \citep[see][for more detailed definitions
  of these samples]{okabe10}. We find that the resulting mean
ellipticity from the faint galaxy catalogue, which is much more
significantly contaminated by the cluster member galaxies, is $\langle
e \rangle=0.44\pm 0.04$.  The result is fully consistent with the
result adopting ``red+blue'' sample, which indicates that the dilution
effect is not very important in constraining cluster ellipticities. 
This is probably because the ellipticity is sensitive mainly to the
two-dimensional shear pattern rather than shear amplitudes which are
more affected by the dilution. Note that profile parameters, cluster
mass and concentration, are sensitive to the shear amplitudes,
therefore are more affected by the dilution \citep{okabe10}.  

\subsection{Effect of subpeaks on the ellipticity 
measurement}\label{sec:subst}

One may argue that the large cluster ellipticity we measured is an
artifact due to the projection effect of foreground/background 
structures or prominent substructures in the cluster region,
which may cause an apparent large elongation in the mass distribution
for the cluster region. To explore this possibility, we visually
inspect mass and luminosity density maps for our sample of clusters
\citep{okabe10}. Among the 18 clusters used for the statistical
analysis of the cluster shape, we find that A267, ZwCl0823, A697, and
A2631 have probable subpeaks which might affect our ellipticity
measurements. We remove these 4 clusters, redo analysis, and find the
mean ellipticity to be $\langle e \rangle=0.48\pm 0.05$, in good
agreement with our main result.  

While the effect of cluster substructures on the cluster shape
measurement turns out to be not so significant compared with the
statistical measurement uncertainty (Appendix~\ref{sec:app}), we also
test this by further removing A68, A209, A291, and A383, which show
large misalignments between the distributions of dark matter and member
galaxies and appear to possess large substructures (see
\S\ref{sec:mem}). The mean ellipticity for the remaining 10 clusters
is $\langle e \rangle=0.46\pm 0.06$, which is still quite consistent
with our main result. Therefore, together with the result of simple
simulations in Appendix~\ref{sec:app}, we conclude that the projection
effect or the effect of cluster substructures are negligible.  

\section{Conclusion}
\label{sec:conclusion}

In this paper, we have presented a systematic study to explore the
ellipticity of cluster mass distribution directly from the 2D pattern
of weak lensing distortion measured from background galaxy images. Our
sample of high quality Subaru/Suprime-cam images for 25 X-ray
selected clusters \citep{okabe10} has enabled reliable statistical
studies of dark matter distributions in massive cluster in great
details. We developed a method of comparing the measured distortion
patter with the prediction computed from an elliptical NFW model,
where the halo ellipticity is modeled in the projected mass density
and the resulting shear field is computed by solving the 2D Poisson
equation (see \S~\ref{sec:model}).  Also notable is our method
properly includes the error covariance matrix arising from the shot
noise due to intrinsic ellipticities as well as the cosmic shear
contamination due to large scale structures that are at different
redshifts, but along the same line of sight to the cluster (see
\S~\ref{sec:2dfit}).  By conducting the 2D shear fitting, we can
extract information on the shape of dark matter distribution in the
cluster region in addition to the cluster mass and the profile
parameters (halo concentration). We have shown several interesting
results, and our findings are summarised as follows. 

From a sample of 18 clusters, we have determined the mean ellipticity
of the dark matter distribution to be $\langle e \rangle=\langle 1-b/a
\rangle =0.46\pm 0.04$ (1$\sigma$), in excellent agreement with the
prediction of the triaxial halo model by \citet{jing02} (see
Figure~\ref{fig:ell}). By comparing the results with the fitting
results obtained from the spherical NFW model ($e=0$), we concluded
that the ellipticity is detected at $7\sigma$ confidence level, which
has been achieved without assuming the location of mass peaks and the
alignment between mass and light. Interestingly, we found that the
constraints on halo ellipticity parameters are not largely degenerated 
with halo mass and concentration, implying that the halo ellipticity
is constrained mainly from the 2D shear pattern rather than the shear
amplitudes (see Figure~\ref{fig:cont_a2390}). 

According to the CDM predictions, dynamically young and more massive
haloes tend to have larger ellipticities. That is, the halo
ellipticity should depend on redshift and mass of clusters as well as
on cosmological models. However, our results are not sufficient to
discriminate the model differences due to the limited size of our
cluster sample (see Figure~\ref{fig:ell_paras}). 

It appears that our results are robust against several possible
systematics, such as the dilution effect, substructures in clusters,
and projections of other haloes along the line-of-sight. For masses
and concentrations, we find no significant systematic differences
between the 1D and 2D analysis (see Figure~\ref{fig:mc_1d2d}). The
mass centroid positions constrained from weak lensing is typically
consistent with the BCG positions with a typical accuracy of $\sim
20$~arcseconds ($\sim 50h^{-1}$~kpc), although there are a few
clusters that show significant offsets (see Figure~\ref{fig:dis}). 
We have found that the correlation between the orientations of the
dark matter distribution and cluster member galaxy distributions is
not very strong, possibly because of the limitation of our member
galaxy sample which is defined photometrically using only two-band
images. Therefore it is of interest to repeat the analysis using
spectroscopically confirmed cluster members.

Overall, our results demonstrate the importance of analysing the
full 2D distributions of cluster weak lensing data. The 2D shear map
contains much richer information than azimuthally-averaged 1D
tangential shear profiles which have been a main focus in most of
previous studies. Direct tests of dark matter distributions should
grow its importance in the era of wide-field imaging surveys which
will bring an exciting opportunity to explore the full use of the
method presented in this paper in order to obtain stringent,
quantitative constraints on the collisionless CDM scenario on small
non-linear scales.  

\section*{Acknowledgments}
We would like to thank Masataka~Fukugita, Hajime Kawahara, Gus Evrard,
Arif Babul, James Taylor, and the
LoCuSS team members for useful discussions and comments.   
We are grateful to Alastair Sanderson for sharing X-ray centroid data
with us. 
We also thank an anonymous referee for many useful suggestions.
This work was initiated during a visit of MO to the Institute for the
Physics and Mathematics of the Universe (IPMU), whose hospitality and
support is gratefully acknowledged. 
This work was supported in part by Department of Energy contract
DE-AC02-76SF00515, and by Grants-in-Aid for Scientific Research from
the JSPS to MT (Nos. 17740129 and 18072001) and to NO
(No. 20740099). GPS acknowledges support from the Royal Society and STFC. 
This work is supported in part by Japan Society for
Promotion of Science (JSPS) Core-to-Core Program ``International
Research Network for Dark Energy'', by Grant-in-Aid for Scientific
Research on Priority Areas No. 467 ``Probing the Dark Energy through
an Extremely Wide \& Deep Survey with Subaru Telescope'', and by World
Premier International Research Center Initiative (WPI Initiative),
MEXT, Japan.  


\appendix

\section{Impact of a subpeak on cluster shape measurements}
\label{sec:app}

\begin{figure}
\begin{center}
 \includegraphics[width=0.9\hsize]{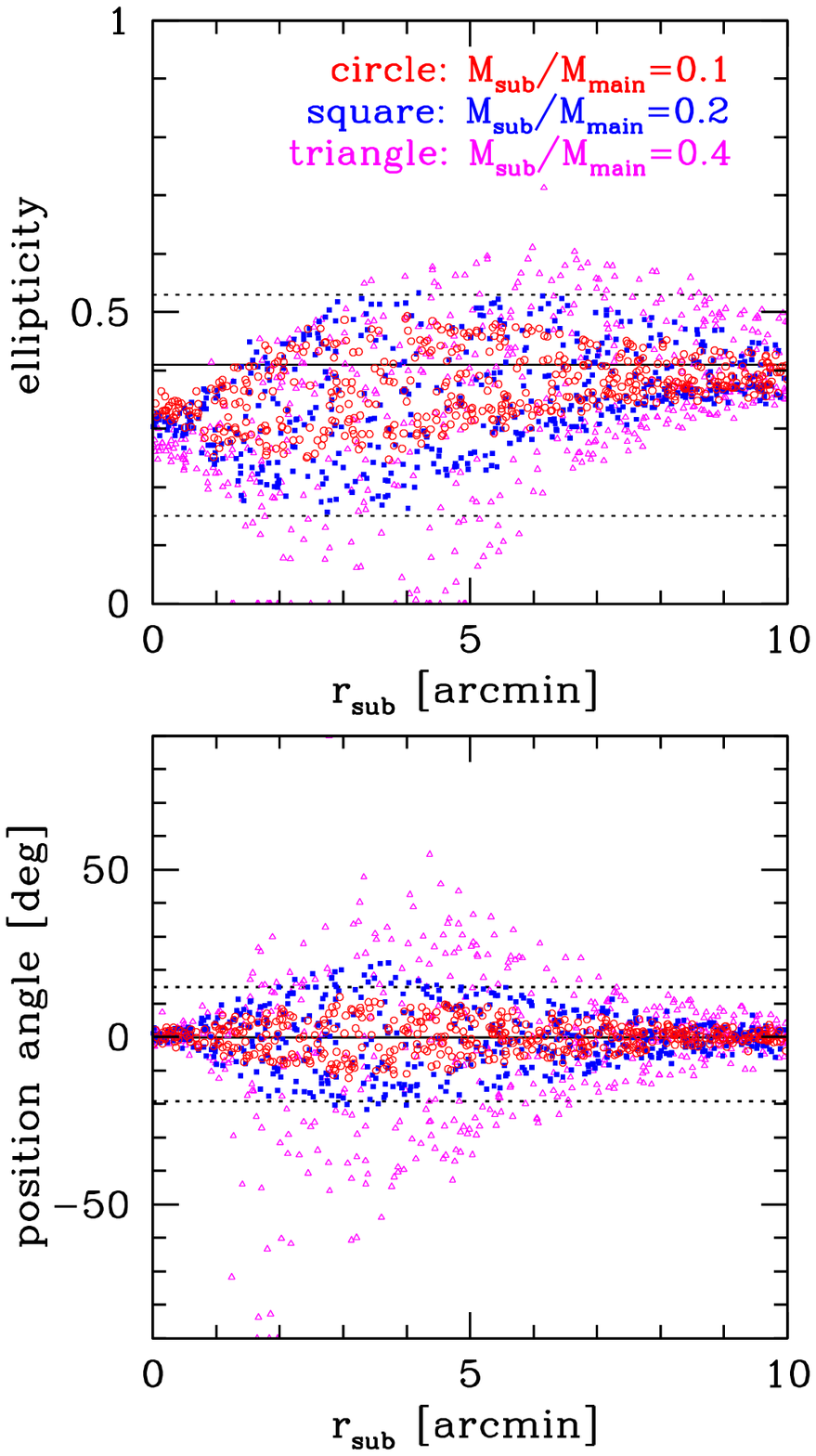}
\end{center}
\caption{The effect of a subpeak on measurements of cluster
  ellipticities ({\it upper}) and position angles ({\it lower}). We
  consider three subpeaks with the mass ratio of $f_{\rm sub}=M_{\rm
    sub}/M_{\rm main}=0.1$ ({\it squares}), $0.2$ ({\it circles}), and
  $0.4$ ({\it triangles}). We study the effect by placing a subpeak
  with its centre randomly chosen, fitting the two-peak system
  assuming a single elliptical NFW component, and deriving best-fit
  parameter values. We repeat this procedure 500 times for each mass
  ratio, and plot best-fit values as a function of the subpeak
  location from the main halo. Horizontal solid and dotted lines
  indicate best-fit values and marginalised $1\sigma$ errors in the
  absence of the subpeak. See text for more details. 
\label{fig:subpeak}}
\end{figure}

We conduct simple simulations to examine the impact of a single 
substructure on cluster shape measurements. We assume a main halo,
modelled by an elliptical NFW profile, with 
$M_{\rm main}=8\times 10^{14}h^{-1}M_\odot$, 
$c_{\rm vir}=6.0$, $e=0.4$, $\theta_e=0$~deg. We then add a spherical
(i.e., $e=0$) NFW subpeak, whose centre is chosen randomly, with its
mass $M_{\rm sub}=f_{\rm sub}M_{\rm main}$. The concentration
parameter of the subpeak taken to be same as that of the main halo. 
The cluster and source redshifts are assumed to $z_l=0.3$ and 
$z_s=1.0$, respectively. Given the mass model, we randomly distribute
source galaxies in the $20'\times20'$ region centred at the main halo
with the number density of $n=15$~arcmin$^{-2}$, and compute reduced
shear for each galaxy.  We then repeat the same fitting procedure used
in this paper (\S\ref{sec:fit}), which assumes a single elliptical NFW
component, to derive the shape and orientation of this compound
system. All the source galaxies have the shape measurement uncertainty
$\sigma_g=0.4$. The effect of large-scale structure along the
line-of-sight is not included in this simulation. 

First we perform the simulation without any subpeak. We find that all
the parameters are recovered well within $1\sigma$ statistical
uncertainties, which validates our fitting procedure using
two-dimensional square grids with a grid size of $1'$. The best-fit
ellipticity and position angles are $e=0.41^{+0.12}_{-0.25}$ and
$\theta_e=-0.1^{+15.1}_{-19.0}$~deg, respectively.   

Next we add a subpeak to see how the best-fit values can be biased. 
We consider three cases, $f_{\rm sub}=0.1$, $0.2$, and $0.4$. For each
case we generate 500 realisations with different subpeak locations. 
Figure~\ref{fig:subpeak} shows the result. We find that the effect of
such a subpeak is modest. The subpeak can have significant (i.e.,
comparable to the statistical uncertainty) impact on
cluster shape measurements only when $f_{\rm sub}>0.2$, which appears
to be quite extreme at least for the case that the subpeak is
physically associated with the main halo. Interestingly, we find that
the subpeak has the largest impact when it is located at $3' \la r \la
5'$ which is the most important region for constraining the cluster
ellipticity (\S\ref{sec:radi}).

\section{Masses and concentration parameters for $\Delta=500$}
\label{sec:m500}

In this paper we have presented results for the virial overdensity
$\Delta_{\rm vir}$. However, different values of the overdensity may
be suited for various applications. For instance, X-ray observables
are more directly connected with masses defined by the overdensity
$\Delta=500$ rather than the virial overdensity
\citep[e.g.,][]{vikhlinin09}. Moreover, \citet{okabe10} showed that
the fractional error on the cluster mass measured from weak lensing is
also minimised for $\Delta\simeq 500$. In Table~\ref{table:m500}, we
show our 2D shear fitting results adopting the overdensity
$\Delta=500$ instead of the fiducial virial overdensity $\Delta_{\rm
 vir}\simeq 110$. Note that constraints on the ellipticity, position
angle, and the mass centroid are unaffected by the adopted value of $\Delta$. 

\begin{table}
\caption{Best-fit masses and concentration parameters for the
  overdensity $\Delta=500$. 
\label{table:m500}}   
\begin{tabular}{@{}ccc}
 \hline
  Name
  & $M_{500}$
  & $c_{500}$ \\
  & ($10^{14}h^{-1}M_\odot$) 
  & \\
\hline
          A68 & $  3.03^{ +1.32}_{ -0.71}$ & $  1.66^{ +2.25}_{ -1.24}$ \\ 
         A115 & $  5.19^{ +0.36}_{ -2.70}$ & $  0.48^{ +0.01}_{ -0.19}$\\ 
         A209 & $  6.42^{ +1.32}_{ -1.48}$ & $  0.97^{ +0.61}_{ -0.32}$\\ 
      RXJ0142 & $  2.75^{ +0.61}_{ -0.50}$ & $  2.73^{ +1.94}_{ -0.68}$ \\ 
         A267 & $  2.04^{ +0.48}_{ -0.35}$ & $  5.53^{ +4.93}_{ -1.89}$\\ 
         A291 & $  2.70^{ +0.88}_{ -0.54}$ & $  1.30^{ +1.40}_{ -0.71}$\\ 
         A383 & $  1.56^{ +0.18}_{ -0.13}$ & $ 31.58^{ +5.52}_{ -4.68}$\\ 
         A521 & $  3.79^{ +1.32}_{ -0.87}$ & $  0.53^{ +0.37}_{ -0.26}$\\ 
         A586 & $  3.53^{ +1.91}_{ -0.90}$ & $  5.16^{ +5.13}_{ -2.72}$ \\ 
     ZwCl0740 & $  6.56^{ +2.36}_{ -2.64}$ & $  0.58^{ +0.47}_{ -0.23}$\\ 
     ZwCl0823 & $  4.82^{ +1.22}_{ -1.44}$ & $  1.23^{ +1.07}_{ -0.60}$ \\ 
        A611  & $  4.46^{ +1.10}_{ -0.81}$ & $  0.69^{ +0.76}_{ -0.36}$ \\ 
         A689 & $  0.98^{ +0.47}_{ -0.54}$ & $  0.08^{ +0.13}_{ -0.03}$\\ 
         A697 & $  6.79^{ +1.32}_{ -1.45}$ & $  0.90^{ +0.63}_{ -0.34}$ \\ 
         A750 & $  3.15^{ +7.11}_{ -1.24}$ & $  0.00^{ +0.00}_{ -0.00}$\\ 
        A1835 & $  6.38^{ +1.35}_{ -1.41}$ & $  1.65^{ +0.60}_{ -0.47}$ \\ 
     ZwCl1454 & $  2.07^{ +0.45}_{ -0.80}$ & $  1.10^{ +1.02}_{ -0.76}$\\ 
     ZwCl1459 & $  2.91^{ +1.11}_{ -0.64}$ & $  4.06^{ +5.03}_{ -1.68}$\\ 
        A2219 & $  4.36^{ +1.27}_{ -0.81}$ & $  5.60^{ +2.68}_{ -2.74}$ \\ 
      RXJ1720 & $  2.26^{ +0.92}_{ -0.52}$ & $  6.32^{ +7.62}_{ -2.50}$\\ 
        A2261 & $  5.74^{ +1.40}_{ -0.87}$ & $  2.86^{ +0.95}_{ -0.88}$\\ 
      RXJ2129 & $  1.42^{ +2.31}_{ -0.56}$ & $  0.00^{ +0.01}_{ -0.00}$\\ 
        A2390 & $  4.56^{ +1.21}_{ -0.85}$ & $  3.26^{ +2.39}_{ -1.22}$ \\ 
        A2485 & $  1.88^{ +0.60}_{ -0.55}$ & $  6.70^{+10.99}_{ -4.10}$  \\ 
        A2631 & $  2.85^{ +0.70}_{ -0.60}$ & $  4.60^{ +6.01}_{ -1.77}$ \\ 
\hline
\end{tabular}
\end{table}

\label{lastpage}

\end{document}